\documentclass[twocolumn,preprintnumbers,aps,pra,superscriptaddress,amsmath,amssymb,balancelastpage]{revtex4-1}

\usepackage{graphicx}
\usepackage{dcolumn}
\usepackage{bm}
\usepackage{bbm}
\usepackage{hyperref}
\usepackage{color}
\usepackage{multirow}
\usepackage{amsmath,amssymb}

\newcommand{\ket}[1]{\ensuremath{\left| #1 \right>}} 
\newcommand{\NV}[1]{\ensuremath{^{#1}}NV} 
\newcommand{\N}[1]{\ensuremath{^{#1}}N} 
\newcommand{\C}[1]{\ensuremath{^{#1}}C} 
\begin{document}

\title{Spin dynamics of diamond nitrogen-vacancy centres at the ground state level anti-crossing and all-optical low frequency magnetic field sensing}

\author{David A. Broadway}
\affiliation{Centre for Quantum Computation and Communication Technology, School of Physics, The University of Melbourne, VIC 3010, Australia}

\author{James D. A. Wood}
\affiliation{Centre for Quantum Computation and Communication Technology, School of Physics, The University of Melbourne, VIC 3010, Australia}

\author{Liam T. Hall}
\affiliation{School of Physics, The University of Melbourne, VIC 3010, Australia}

\author{Alastair Stacey}
\affiliation{Centre for Quantum Computation and Communication Technology, School of Physics, The University of Melbourne, VIC 3010, Australia}
\affiliation{Element Six Innovation, Fermi Avenue, Harwell Oxford, Didcot, Oxfordshire OX110QR, United Kingdom}

\author{Matthew Markham}
\affiliation{Element Six Innovation, Fermi Avenue, Harwell Oxford, Didcot, Oxfordshire OX110QR, United Kingdom}

\author{David A. Simpson}
\affiliation{School of Physics, The University of Melbourne, VIC 3010, Australia}	

\author{Jean-Philippe Tetienne}\email{jtetienne@unimelb.edu.au}
\affiliation{Centre for Quantum Computation and Communication Technology, School of Physics, The University of Melbourne, VIC 3010, Australia}

\author{Lloyd C. L. Hollenberg}
\affiliation{Centre for Quantum Computation and Communication Technology, School of Physics, The University of Melbourne, VIC 3010, Australia}
\affiliation{School of Physics, The University of Melbourne, VIC 3010, Australia}	

\date{\today}

\begin{abstract}
We investigate the photo-induced spin dynamics of single nitrogen-vacancy (NV) centres in diamond near the electronic ground state level anti-crossing (GSLAC), which occurs at an axial magnetic field around 1024 G. Using optically detected magnetic resonance spectroscopy, we first find that the electron spin transition frequency can be tuned down to 100 kHz for the \NV{14} centre, while for the \NV{15} centre the transition strength vanishes for frequencies below about 2 MHz owing to the GSLAC level structure. Using optical pulses to prepare and readout the spin state, we observe coherent spin oscillations at 1024 G for the \NV{14}, which originate from spin mixing induced by residual transverse magnetic fields. This effect is responsible for limiting the smallest observable transition frequency, which can span two orders of magnitude from 100 kHz to tens of MHz depending on the local magnetic noise. A similar feature is observed for the \NV{15} centre at 1024 G. As an application of these findings, we demonstrate all-optical detection and spectroscopy of externally-generated fluctuating magnetic fields at frequencies from 8 MHz down to 500 kHz, using a \NV{14} centre. Since the Larmor frequency of most nuclear spin species lies within this frequency range near the GSLAC, these results pave the way towards all-optical, nanoscale nuclear magnetic resonance spectroscopy, using longitudinal spin cross-relaxation.
\end{abstract}	

\maketitle

\section{Introduction}

Detection and identification of spin species using established techniques such as magnetic resonance spectroscopy proves to have a host of applications in materials science, chemistry and biology. However, these techniques are limited in sensitivity, and thus require macroscopic ensembles of spins in order to produce a measurable signal \cite{Blank2003}. A variety of techniques have been developed over the last decade to extend magnetic resonance spectroscopy to the nanometre scale \cite{Poggio2010,Artzi2015,Bienfait2016,Rondin2014,Schirhagl2014}. Notably, methods based on the nitrogen-vacancy (NV) centre in diamond \cite{Doherty2013} have attracted enormous interest owing to their ability to operate under conditions compatible with biological samples \cite{Schirhagl2014,McGuinness2011}. While significant progress has been made with NV-based sensing in the last few years \cite{Rondin2014}, spectroscopy at the single nuclear spin level remains a major challenge. 

The detection of external spins with the NV centre is generally achieved through measuring the longitudinal spin relaxation rate ($T_1$ processes) \cite{Steinert2013,Tetienne2013,Kaufmann2013,Ermakova2013,Sushkov2014} or transverse spin relaxation rate (dephasing, or $T_2$ processes) \cite{Ermakova2013,Maze2008,DeLange2011,McGuinness2013} of the NV's electron spin, as they are sensitive to the magnetic field fluctuations produced by the target spins \cite{Cole2009,Hall2009,Laraoui2010}. To obtain spectral information on the target spins, most studies so far have focussed on using $T_2$-based techniques, which have been applied to the spectroscopy of small ensembles of either electronic \cite{Grotz2011,Laraoui2012,Mamin2012,Knowles2013} or nuclear spins \cite{Mamin2013,Staudacher2013,Loretz2014,Muller2014,DeVience2015}. However, spectroscopy can also be achieved by relying on $T_1$ processes, via cross-relaxation between a probe spin (the NV centre's electron spin) and the target spins \cite{Jarmola2012,Wang2014,VanderSar2015,Hall2016,Wood2016}. The $T_1$-based approach to spectroscopy introduced in Ref. \cite{Hall2016} allows for nanoscale, all-optical, wide-band magnetic resonance spectroscopy \cite{Wood2016}. As such, it represents a promising alternative to $T_2$-based approaches, as the latter require radiofrequency (RF) driving of the probe and/or the target, which poses various technical challenges in addition to limiting the accessible frequency range. 

In cross-relaxation spectroscopy, the $T_1$ of the NV spin is monitored while varying the strength of an applied axial magnetic field, $B_z$ (Fig. \ref{Fig:intro}a). When the transitional energy between two of the NV eigenstates (generally \ket{0_e} and \ket{-1_e}, where the number refers to the electron spin projection $m_e$) is equal to that of a target spin, cross-relaxation occurs, which results in an increase in the relaxation rate, $1/T_1$, of the NV \cite{Hall2016,Wood2016}. This increase can be measured by purely optical means, even for a single NV centre \cite{Tetienne2013,Kaufmann2013,Ermakova2013,Sushkov2014}. By scanning across a range of magnetic field strengths, a resonance spectrum of the target spins can be obtained, which can be deconvolved to produce the target spin spectrum \cite{Hall2016}. This technique has been recently used to measure electronic spin resonance (ESR) spectra of P1 centres within the diamond at magnetic fields of 460-560 G, corresponding to transition frequencies of 1300-1600 MHz \cite{Hall2016,Wood2016}. In order to detect nuclear magnetic resonances (NMR), the NV transition frequency must be matched to the Larmor frequency of the target nuclear spins, which is generally of order a few MHz. This occurs when the states \ket{0_e} and \ket{-1_e} approach degeneracy, at a magnetic field $B_z\approx1024$~G (Fig. \ref{Fig:intro}b). However, in this region the NV experiences a complex Ground State Level Anti-Crossing (GSLAC, Figs. \ref{Fig:intro}c,d) \cite{He1993} due to hyperfine interaction of the NV electron spin with its own nuclear spin ($^{14}$N or $^{15}$N), which also has a coupling strength of several MHz. The hyperfine interaction causes spin mixing, which may prevent the NV electron spin from being initialised and read out \cite{Epstein2005}. Therefore, a detailed understanding of the spin dynamics at the GSLAC is required in order to assess the potential for performing $T_1$-based spectroscopy of nuclear spins as initially explored in Ref. \cite{Wood2016}. While the GSLAC of the NV centre has been previously studied and exploited in several works \cite{He1993,Wei1999,Wilson2003,Epstein2005,Fuchs2011,Wang2013,Wang2015a,Wickenbrock2016}, there is little knowledge about how $T_1$ varies and how the spin dynamics (including optical initialization) behaves at transition frequencies relevant to NMR, close to the GSLAC.   

In this paper, we investigate the spin dynamics at the GSLAC for both \NV{14} and \NV{15} spin systems in various diamond samples. We begin by looking at the computed energy spectra of both spin systems and compare them to optically detected magnetic resonance (ODMR) measurements at their respective GSLACs. The short time spin dynamics of the \NV{14} spin at the GSLAC are probed using optical pulses, revealing a feature at $B_z\approx1024$ G which manifests itself in either coherent spin oscillations, or in a simple polarisation drop depending on the sample. This feature is explained by spin mixing induced by residual transverse magnetic fields. The \NV{15} centre shows a similar feature at $B_z\approx 1024$ G. In addition to these narrow features, we find that the spin  polarisation and $T_1$ time remains constant across the GSLAC, implying that the NV centre can be used to detect magnetic signals at low frequencies via $T_1$ measurements. Finally, we demonstrate one such application by performing all-optical spectroscopy of fluctuating magnetic fields generated at known frequencies, mimicking those of nuclear spins. This suggests that it is possible to perform NMR spectroscopy via longitudinal cross-relaxation near the GSLAC.  

\section{Energy levels of the NV centre at the GSLAC}

\begin{figure}
\includegraphics[width=0.45\textwidth]{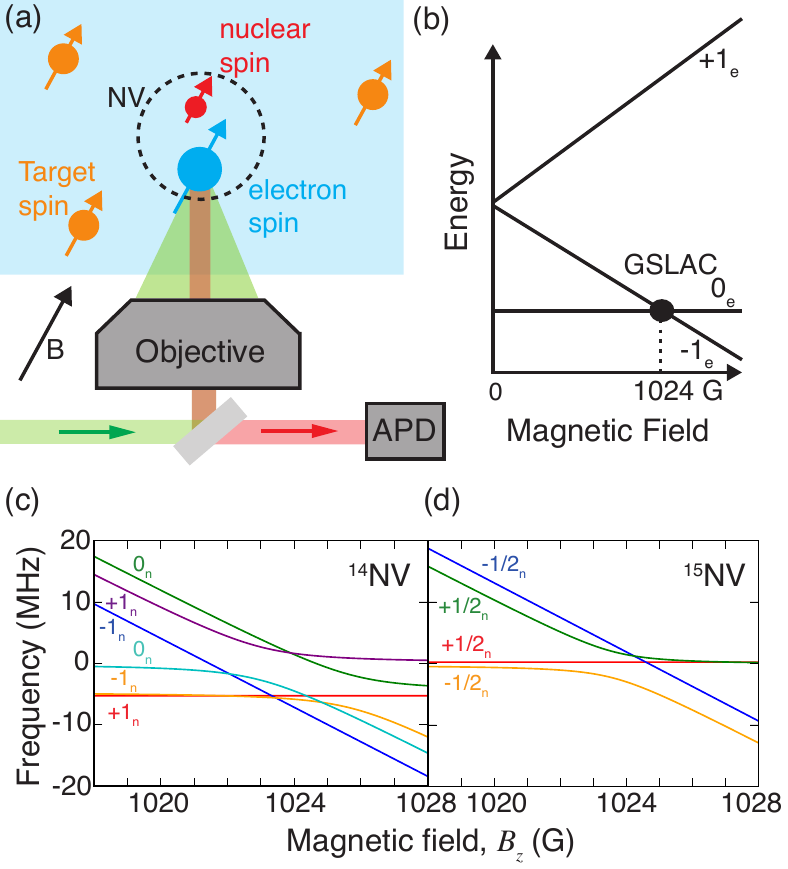}
\caption{ (a) Schematic view of the system under study: the NV centre in diamond comprises a nuclear spin and an electron spin, which can be initialised and read out optically using a confocal microscope equipped with an avalanche photodiode (APD); it is surrounded by target spins located within the diamond or external to it. (b) Energy structure of the NV electronic ground state, showing the Zeeman splitting under a magnetic field applied along the NV symmetry axis. The three electronic spin levels are labelled by their spin projection $m_e$. (c,d) Calculated hyperfine structure for the \NV{14} (c) and \NV{15} centre (d), plotted near the GSLAC where the two branches $m_e=0$ and $m_e=-1$ cross. The different levels are labelled by the nuclear spin projection $m_n$ of the unperturbed states (away from the GSLAC). Due to hyperfine interaction, some of the levels exhibit an avoided crossing at the GSLAC.}
\label{Fig:intro}
\end{figure}

The NV spin system consists of a nitrogen atom adjacent to a vacancy in the carbon lattice of diamond. It comprises a pair of electrons (forming a spin-1) and a nuclear spin, which is a spin-1 for \NV{14} and spin-1/2 for \NV{15} (Fig. \ref{Fig:intro}a). Due to spin-spin interaction, the electronic spin states \ket{\pm 1_e} are split from \ket{0_e} by $D/2\pi\approx2.87$ GHz. The degeneracy of the \ket{\pm 1_e} can be lifted by the application of an external magnetic field along the NV centre's symmetry axis, defined as the $z$ axis, as shown in Fig. \ref{Fig:intro}b. The \ket{-1_e} and \ket{0_e} states cross at a field around $B_z=D/\gamma_e\approx 1024$ G (where $\gamma_e$ is the electron gyromagnetic ratio), shown as a black dot in Fig. \ref{Fig:intro}b. However, hyperfine interaction with the nitrogen nuclear spin causes an avoided crossing, the GSLAC, which is the main focus of this paper.

The energy spectrum of the NV electronic ground state near the GSLAC can be found by solving for the eigenvalues of the spin Hamiltonian. The relevant Hamiltonians for the \NV{14} and \NV{15} cases, expressed in units of angular frequencies, are
\begin{align}
{\cal H}(\text{\NV{14}}) &= D S_z^2 + \gamma_e B_z S_z - \gamma_n B_z I_z + Q I_z^2 \\
&\qquad \qquad + A_{\parallel} S_z I_z + A_{\perp}\left( S_x I_x + S_y I_y  \right), \notag\\
{\cal H}(\text{\NV{15}}) &= D S_z^2 + \gamma_e B_z S_z - \gamma'_n B_z I'_z \\
&\qquad\qquad + A_{\parallel}^\prime S_z I'_z + A_{\perp}^\prime \left( S_x I'_x + S_y I'_y  \right), \notag 
\end{align}
where ${\bf S}=(S_x,S_y,S_z)$ is the electron spin operator, {\bf I} is the nuclear spin operator, and $\gamma_n$ is the nuclear gyromagnetic ratio. The magnetic field is aligned along the NV axis, with strength $B_z$. The primed symbols refer to the \NV{15} case. The longitudinal and transverse hyperfine parameters are denoted as $A_\parallel$ and $A_\perp$, whose values are $A_\parallel/2\pi = -2.14$ MHz and $A_\perp/2\pi = -2.7$ MHz for \NV{14}, $A_\parallel^\prime/2\pi = 3.03$ MHz and $A_\perp^\prime/2\pi = 3.65$ MHz for \NV{15} \cite{Felton2009}. In addition, the \NV{14} has a quadrupole coupling with strength $Q/2\pi = -5.01$ MHz \cite{Felton2009}.

The energy spectrum is obtained by evaluating the eigenvalues of the Hamiltonian at different axial field strengths $B_z$ and is shown in Fig.~\ref{Fig:intro}c for \NV{14} and Fig.~\ref{Fig:intro}d for \NV{15}. The manifold associated with the spin projection \ket{+1_e} is not shown as it lies about 6 GHz above the manifold spanned by \ket{0_e} and \ket{-1_e} and does not contribute to the effects discussed in this paper. We thus consider only the 6 lower-energy states for \NV{14}, and the 4 lower states for \NV{15}. Away from the GSLAC, the eigenstates have well-defined spin projections along the $z$ axis. In the electron-nuclear spin space, we denote states as \ket{m_e,m_n} where $m_e$ ($m_n$) is the electronic (nuclear) spin projection along the $z$ axis. The hyperfine coupling between the NV nuclear and electron spins results in a splitting of the nuclear spin states for each electronic spin state, due to the longitudinal component $A_\parallel$. Near the GSLAC, the perpendicular component of the hyperfine ($A_\perp$) induces a mixing of some of the $z$-basis states. This effect is noticeable when the quantization energy between said states becomes of order $A_\perp$. In the case of the \NV{14}, the states \ket{0, +1} and \ket{-1, -1} do not mix as they have no hyperfine coupling to any other state, and as such, they remain eigenstates. The rest of the NV states are mixed, that is, the eigenstates are superpositions of $z$-basis states, creating an avoided crossing. The \NV{15} spin also exhibits mixing at the GSLAC. In particular the \ket{0,-1/2} and \ket{-1,+1/2} states become mixed while the \ket{0,+1/2} and \ket{-1,-1/2} states remain eigenstates.

\section{Optically detected magnetic resonance at the GSLAC} \label{Sec:ODMR}

\begin{figure}
\includegraphics[width=0.47\textwidth]{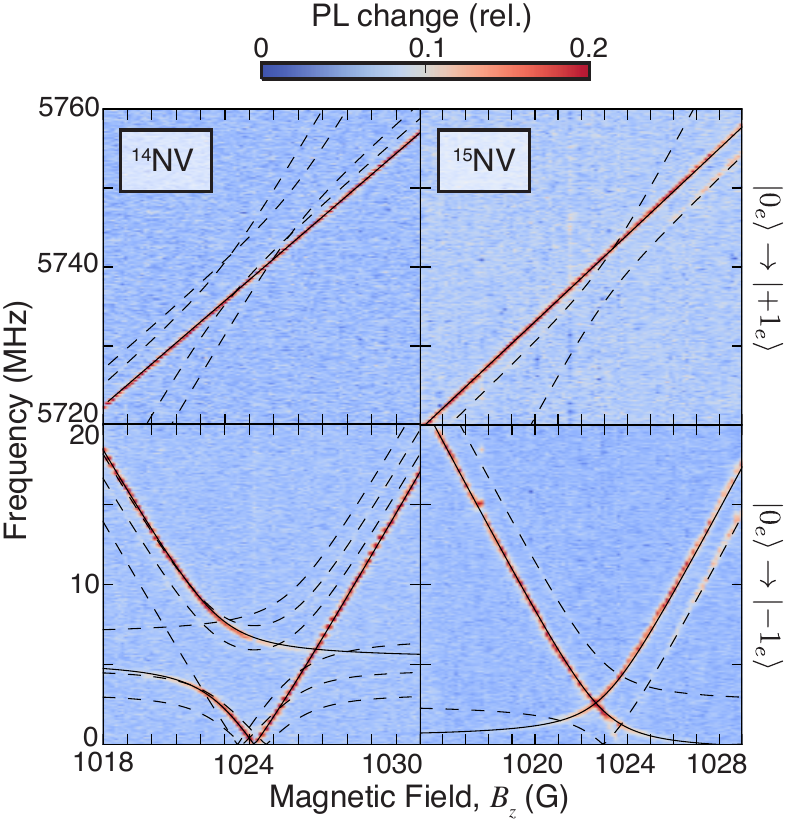}
\caption{ODMR spectra of a \NV{14} (left panels) and \NV{15} (right panels) centre, measured as a function of the axial magnetic field strength near the GSLAC. The top panels show the electron spin transitions $|0_e\rangle\rightarrow|+1_e\rangle$, while the bottom panels show $|0_e\rangle\rightarrow|-1_e\rangle$. Overlaid on the graph are the theoretical frequencies of all allowed or partly allowed (via spin mixing) transitions. However, dynamic nuclear spin polarisation makes some of the transitions dominant (shown as solid lines), the other transitions having comparatively small or vanishing contrast (dashed lines).} 
\label{Fig:ODMR}
\end{figure}

Experimentally, one can probe the NV energy spectrum using optically detected magnetic resonance (ODMR) spectroscopy \cite{Gruber1997}. This is achieved by measuring the photoluminescence (PL) intensity of the NV centre while varying the frequency of an applied RF field, using a purpose-built confocal microscope with green laser excitation (Fig. \ref{Fig:intro}a). The laser serves both to initialize the NV in the electronic spin state \ket{0_e}, and read out the spin state following an RF pulse, exploiting the fact that \ket{\pm1_e} emit less PL on average than \ket{0_e}  \cite{Manson2006}. Thus, ODMR allows us to probe the electron spin transitions $|0_e\rangle\rightarrow|\pm 1_e\rangle$. We recorded ODMR spectra for magnetic fields varied from 1018 G to 1032 G, for single \NV{14} and \NV{15} centres in a high-purity diamond grown by chemical vapour deposition (CVD). The results are shown in Fig. \ref{Fig:ODMR}, where the top (bottom) panels show the \ket{0_e}$\rightarrow$ \ket{+1_e} (\ket{0_e}$\rightarrow$ \ket{-1_e}) transitions. The left (right) panels correspond to a \NV{14} (\NV{15}) centre. Comparing with the theoretical expectations for the allowed transitions (shown as black lines), it can be seen that only a limited number of expected transitions are observed experimentally. This is because the nuclear spin is efficiently polarised under optical pumping near the GSLAC, owing to hyperfine-induced spin mixing. This effect has been well documented at the excited state level anti-crossing (ESLAC, \cite{Jacques2009,Ivady2015}), but has not been previously quantified experimentally across the GSLAC \cite{Fuchs2011}. Using the relative strengths of the \ket{0_e}$\rightarrow$\ket{+1_e} transitions (top panels in Fig. \ref{Fig:ODMR}), we find that the nuclear spin is polarised to $>90\%$ into \ket{+1_n} for \NV{14} across the whole range of fields scanned here. For the \NV{15} investigated in Fig. \ref{Fig:ODMR}, the nuclear spin is polarised to $>90\%$ in \ket{+1/2_n} up to $B_z\approx 1026$ G, but becomes completely unpolarised above $B_z\approx 1028$ G. 

We now discuss the ODMR spectrum of the \ket{0_e}$\rightarrow$\ket{-1_e} transition (bottom panels in Fig. \ref{Fig:ODMR}). For \NV{14}, the NV is efficiently polarised in the state \ket{0,+1}, which remains an eigenstate at all fields. Away from the GSLAC, the only transition allowed is to the state with the same nuclear spin projection, \ket{-1,+1}. However, near the GSLAC the state \ket{-1,+1} becomes mixed with \ket{0,0}, which creates two eigenstates of the form $\ket{\alpha_\pm} = \alpha_{1\pm}\ket{-1,+1} + \alpha_{2\pm}\ket{0,0}$. This results in an avoided crossing feature centred at $B_z\approx 1022$ G, around the transition frequency $\omega\approx 5$ MHz which corresponds to the quadrupole coupling $Q$. This leads to little mixing at the allowed crossing at $B_z\approx1024$ G, which means the dominant transition here is \ket{0,+1}$\rightarrow$\ket{-1,+1}. We observed clear ODMR signatures with resonance frequencies down to 100 kHz in this sample. This implies that the NV spin could be resonantly coupled to most nuclear spin species, which have Larmor frequencies ranging typically from 500 kHz to 5 MHz at this magnetic field. However, the minimum observable transition frequency was found to be highly sample dependent, as will be discussed in section \ref{Sec:Dynamics}. 

On the other hand, the \NV{15} exhibits a very different spectrum near the GSLAC (Fig. \ref{Fig:ODMR}, right-hand panels). Under optical pumping, it is efficiently polarised in the state \ket{0,+1/2}, which remains an eigenstate at all fields. The states it can transit to under RF driving are superpositions $\ket{\beta_\pm} = \beta_{1\pm}\ket{0,-1/2} + \beta_{2\pm}\ket{-1,+1/2}$, where the avoided crossing is centred approximately around the initial state \ket{0,+1/2} (see Fig. \ref{Fig:intro}d). As a result, the ODMR plot shows two transitions that bend upon approaching vanishing frequencies. They cross at a field $B_z\approx 1024$ G and a transition frequency $\omega'_{\times}/2\pi\approx2.65(2)$ MHz given by
\begin{align}
\omega'_{\times} = \sqrt{\frac{A_\perp^{\prime 2}}{2}+\left(\frac{\gamma'_n}{\gamma_e-\gamma'_n}\right)^2\left(D-\frac{A'_\parallel}{2}\right)^2}.
\end{align}
Incidentally, this allows the perpendicular component of the hyperfine interaction ($A'_\perp$) to be measured directly on a single \NV{15} centre, which gives here $A'_{\perp}/2\pi = 3.69(3)$ MHz, in excellent agreement with the ensemble-averaged value of 3.65(3) MHz reported in Ref. \cite{Felton2009}. The peculiar GSLAC structure of the \NV{15} centre has important consequences for sensing. In particular, the contrast of the transitions decreases rapidly for frequencies below $\omega'_{\times}$ as they become forbidden. As a result, the \NV{15} is unsuited to detecting resonances below about 2 MHz under typical conditions. Most nuclear spin species have transitions within this range, with an exception being hydrogen ($^1$H), which has a Larmor frequency of about 4.4 MHz at this field and could be in principle detected via cross-relaxation with an \NV{15}. 

\section{Photo-induced spin dynamics at the GSLAC} \label{Sec:Dynamics}
      
\begin{figure}
\includegraphics[width=0.47\textwidth]{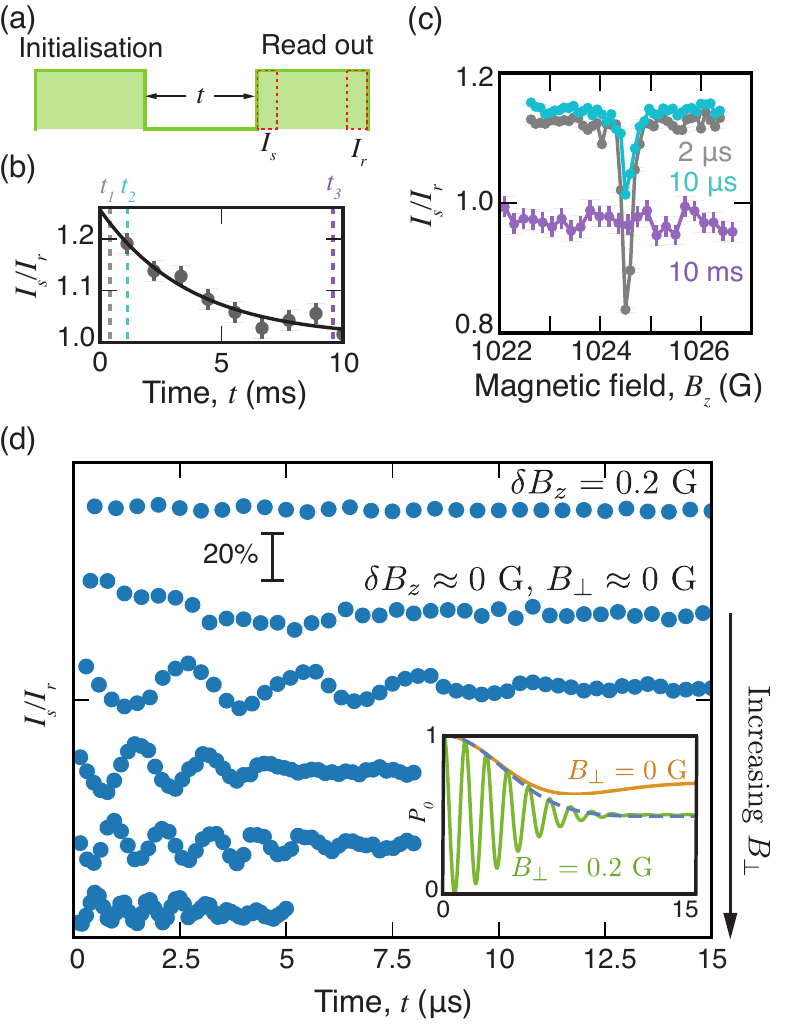}
\caption{(a) Depiction of the laser pulse sequence used to measure the spontaneous spin dynamics, containing an initialisation pulse and a readout pulse after a wait time $t$. (b) Time trace measured for a single \NV{14} centre at a field $B_z=1000$ G. Here a decay with a characteristic time $T_1\approx5$ ms is observed, associated with phonon relaxation. (c) Set of PL scans as a function $B_z$ for various wait times $t$,  under a small residual transverse field $B_\perp$, showing a narrow feature at $B_z\approx1024$ G. The wait times $t=2$ $\mu$s (grey dots), 10 $\mu$s (light blue) and 1 ms (purple) are indicated as dashed lines in (b) with exaggerated positions for ease of viewing. (d) Time traces recorded at $B_z\approx1024$ G. The top curve was measured with a detuning $\delta B_z=0.2$ G, showing no variation in the range $t=0-15~\mu$s. The other curves are recorded with no detuning, but with a transverse magnetic field $B_\perp$ increasing from $\approx0$ G to $\approx0.4$ G (top down). The curves are vertically offset from each other for clarity. Inset: Averaged time traces $\langle P_0(t)\rangle$ computed from Eq. (\ref{Eq:osc}) where $\delta B_z$ and $B_\perp$ are normally distributed, with means $\langle\delta B_z\rangle=0$ and $\langle B_\perp\rangle=0$ G (orange curve) or 0.2 G (green curve) and variances $\sigma_{B_z}=1.3~\mu$T and $\sigma_{B_\perp}=\sqrt{3/4}\sigma_{B_z}$. The dashed blue line is the approximate envelope $e^{-(t/T_\times)^2}$ with $T_\times=(\gamma_e\sigma_{B_\perp})^{-1}=5~\mu$s.}
\label{Fig:C12}
\end{figure}
	
We now investigate the spin population dynamics near the GSLAC. Our aim is to assess the possibility of measuring the longitudinal spin relaxation time ($T_1$), as required in order to perform cross-relaxation spectroscopy and detect nearby nuclear spins \cite{Wood2016}. The $T_1$ time is typically measured by using laser pulses to initialise the NV into \ket{0_e}, and read out the remaining population of \ket{0_e} after a variable delay $t$ (Fig. \ref{Fig:C12}a). In practice, the PL signal is integrated at the start of the readout pulse ($I_s$) and normalised by the PL from the back of the pulse ($I_r$). The normalised signal $I_s/I_r$ can be expressed as \cite{Manson2006} 
\begin{align}
\frac{I_s}{I_r}(t)=a+bP_0(t)
\end{align}
where $P_0(t)=|\langle 0_e|\psi(t)\rangle|^2$ is the population in \ket{0_e} of the current spin state $|\psi(t)\rangle$, and $a\approx 1$ and $b\approx 0.3$ are constants. The resulting time trace $\frac{I_s}{I_r}(t)$ therefore allows us to estimate the initial population $P_0(0)$, which approaches unity under normal conditions \cite{Manson2006}, as well as the evolution of the spin state in the dark. In general (away from the GSLAC or ESLAC), the spin population exhibits a simple exponential decay towards a thermal mixture, i.e. $P_0(t)=\frac{1}{3}+\frac{2}{3}e^{-t/T_1}$ assuming perfect initialisation (see an example in Fig. \ref{Fig:C12}b). In the following, we measure $\frac{I_s}{I_r}(t)$ for different wait times $t$, as a function of the axial magnetic field $B_z$.

\subsection{\NV{14} centres}

We first performed measurements of \NV{14} centres in a ultra-high purity CVD-grown diamond with isotopically purified carbon content ([\C{12}]$>99.99\%$). In this sample, the main source of magnetic noise comes from the bath of remaining \C{13} impurities \cite{Balasubramanian2009}. A scan across the GSLAC for a representative \NV{14} centre is shown in Fig.~\ref{Fig:C12}c, where we probed three time points $t=2$~$\mu$s, 10~$\mu$s, and 10~ms. When the NV is far from the GSLAC crossing, an exponential decay is observed as shown in the full time trace in Fig.~\ref{Fig:C12}b. This decay corresponds to phonon-induced relaxation, with a characteristic time $T_1\approx5$ ms \cite{Jarmola2012}. At the crossing at $B_z\approx1024$ G, however, a sharp variation in signal is observed (Fig. \ref{Fig:C12}c). Here, the NV spin undergoes population oscillations, as indicated by the 2 $\mu$s time point dropping below the 10~ms point. 

To understand this oscillation, we look at the reduced Hamiltonian, ${\cal H}_R$, in the basis that contains the states that cross, $\{|0,+1\rangle,|-1,+1\rangle\}$. In a magnetic field ${\bf B}=(B_x,B_y,B_z)$, this Hamiltonian is given by
\begin{align}
{\cal H}_R =
\begin{pmatrix}
0 & \gamma_e B_\perp\frac{e^{-i\theta}}{\sqrt{2}} \\
\gamma_e B_\perp\frac{e^{+i\theta}}{\sqrt{2}} & \gamma_e\delta B_z
\end{pmatrix} ,
\end{align}
where we introduced the longitudinal detuning from the crossing, $\gamma_e \delta B_z=D-\gamma_e B_z-A_\parallel$, the transverse magnetic field, $B_\perp=\sqrt{B_x^2+B_y^2}$, and the angle $\theta$ defined by $\tan\theta=B_y/B_x$. The transverse field causes a mixing between $|0,+1\rangle$ and $|-1,+1\rangle$ and opens an energy gap associated with a level avoided crossing. Assuming that optical pumping always initialises the NV in the \ket{0,+1} state, and reads out the population in that same state, we then expect oscillations between \ket{0,+1} and \ket{-1,+1} that are mirrored in the PL when in the presence of a transverse magnetic field. Under these conditions, the probability of occupying the state \ket{0,+1} after a wait time $t$ following initialisation is given by
\begin{align}\label{Eq:osc}
P_0(t) = \frac{\delta B_z^2 + B_\perp^2 \left[1 + \cos \left(\gamma_e t \sqrt{\delta B_z^2 + 2 B_\perp^2}\right)\right]}{\delta B_z^2 + 2 B_\perp^2}.
\end{align}
The amplitude of the oscillation vanishes when the detuning is much larger than the transverse field ($\delta B_z\gg B_\perp$), far from the avoided crossing region. This is illustrated in Fig. \ref{Fig:C12}d (top curve), which was recorded with a detuning $\delta B_z\approx0.2$ G $\gg B_\perp$.  On the other hand, near the avoided crossing where the amplitude is maximal, the frequency of the oscillation is expected to increase as $B_\perp$ is increased. This effect was tested through a series of measurements with varying transverse fields. Experimentally this involved using a permanent magnet to align the field at the 1024 G crossing so that no oscillations are detected. The permanent magnet is then moved in the transverse direction ($x$ or $y$) to add a transverse field. The results are shown in Fig. \ref{Fig:C12}d where $B_\perp$ is increased from top down, resulting in faster oscillations. 

Damping of the oscillations is attributed to noise in the local magnetic field. In this sample, the noise comes predominantly from the bath of \C{13} impurities. Examination of this interaction via the rotating wave approximation shows that only the $x$-$z$, $y$-$z$ and $z$-$z$ components of the dipole-dipole coupling to the NV spin need be considered \cite{Hall2014}. As such, the effective magnetic noise from the environment may be regarded as static over the short wait times, $t$, considered here. We assume that the field components $\delta B_z$ and $B_\perp$ are normally distributed with means $\langle\delta B_z\rangle$ and $\langle B_\perp\rangle$ and variances $\sigma^2_{B_z}$ and $\sigma^2_{B_\perp}$, respectively. Averaging Eq. (\ref{Eq:osc}) over these distributions, we find numerically that the decay envelope of $\langle P_0(t)\rangle$ is well approximated by a gaussian function $e^{-(t/T_\times)^2}$, where the characteristic time $T_\times$ is given by $T_\times^{-1}=\gamma_e\sigma_{B_\perp}$, regardless of the means $\langle\delta B_z\rangle$ and $\langle B_\perp\rangle$ (see inset in Fig. \ref{Fig:C12}d). In other words, the damping of the 1024 G oscillations is mainly caused by the fluctuations in the transverse magnetic field. It is interesting to link this damping time $T_\times$ to the dephasing time $T_2^*$ measured in a free induction decay (FID) experiment \cite{Hall2014,Maze2012}. Under the same assumptions, the FID envelope takes the form $e^{-(t/T_2^*)^2}$ where $(T_2^*)^{-1}=\gamma_e\sigma_{B_z}/\sqrt{2}$. Moreover, a bath of randomly placed spins around the NV centre leads to $\sigma^2_{B_\perp}\approx\frac{3}{4}\sigma^2_{B_z}$ on average \cite{Hall2014}, which gives the relation 
\begin{align}
T_\times \approx T_2^*\sqrt{\frac{2}{3}}.
\end{align}
For the NV centre studied in Fig. \ref{Fig:C12}, the damping time of the 1024 G oscillations is $T_\times\approx5-10~\mu$s, estimated from the curves shown in Fig. \ref{Fig:C12}d, hence $\sqrt{\frac{3}{2}}T_\times\approx6-12~\mu$s. This is significantly shorter than the dephasing time $T_2^*>50~\mu$s. We attribute the discrepancy mainly to drifts in the magnetic field applied during the measurements, which leads to overestimating the damping rate $1/T_\times$.

We now consider the case where the magnetic field is aligned along the NV axis, i.e. $\langle B_\perp\rangle=0$. At the crossing when $\delta B_z = 0$, the averaged population $\langle P_0(t)\rangle$ does not oscillate but still decays with a characteristic time $T_\times$ (see inset in Fig. \ref{Fig:C12}d). However, the amplitude of the decay decreases as $\delta B_z$ is increased. We define the width of the 1024 G feature, denoted $\Delta B_z^\times$, as twice the detuning $\delta B_z$ to apply to obtain a maximum population drop of 20\%. We find numerically that $\Delta B_z^\times\approx 4\sigma_{B_\perp}$, which can also be expressed as a function of $T_2^*$ according to
\begin{align} \label{Eq:width} 
\gamma_e\Delta B_z^\times\approx \frac{4}{T_\times}\approx \frac{2\sqrt{6}}{T_2^*}.
\end{align}        
For the NV studied here, we predict a width $\Delta B_z^\times<1~\mu$T for a perfectly aligned background field. We note however that in the measurements of Fig. \ref{Fig:C12}c, the width is instead given by the residual transverse field ($B_\perp\approx0.3$ G in Fig. \ref{Fig:C12}c), which could not be maintained to significantly smaller values for extended periods of time due to drifts in the applied magnetic field. 

The observation of coherent oscillations at the GSLAC suggests a direct application to DC magnetometry. Indeed, the frequency of the oscillation is directly proportional to the strength of the transverse field according to Eq. (\ref{Eq:osc}), assuming $\delta B_z\ll B_\perp$. For photon shot noise limited measurements, the magnetic sensitivity is similar to that obtained by FID measurements \cite{Balasubramanian2009}, with the advantage of being an all-optical technique (no microwave or RF field is required).

\begin{figure}
\includegraphics[width=0.47\textwidth]{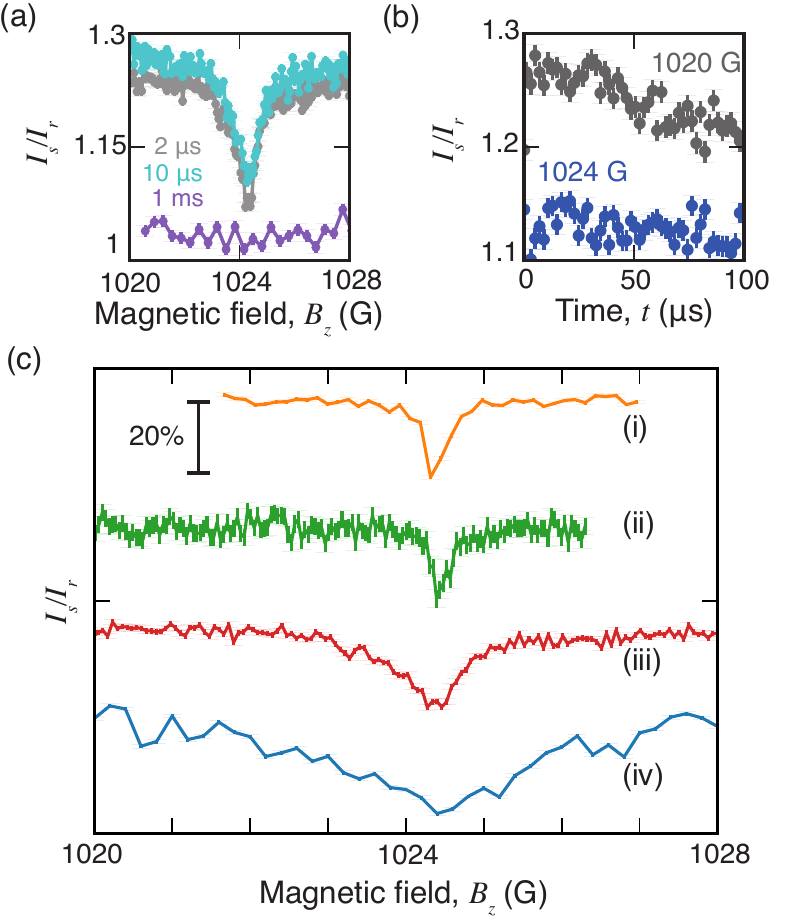}
\caption{(a) PL scan across the GSLAC for a shallow \NV{14} centre in a CVD diamond, with wait times $t=2~\mu$s, $10~\mu$s and 1 ms. Here no oscillation is detected but an overall decrease in spin population is observed at $B_z\approx1024$ G. (b) Time traces recorded at the crossing feature ($B_z\approx1024$ G, blue dots) and away from it ($B_z\approx1020$ G, grey dots). (c) PL scans across the GSLAC recorded with \NV{14} centres in various diamond samples: (i) deep NV in isotopically purified CVD diamond as in Fig. \ref{Fig:C12}, (ii) deep NV in natural isotopic content CVD diamond, (iii) shallow NV in CVD diamond as in (a,b), (iv) deep NV in type-Ib diamond. The curves are vertically offset from each other for clarity.}
\label{Fig:GSLACcomp}
\end{figure}
  
We now compare the spin dynamics of \NV{14} centres at the GSLAC in different diamond samples. Of particular relevance to sensing applications are NV centres implanted close to the diamond surface. We performed measurements of shallow NV centres in a CVD-grown diamond with natural isotopic concentration ([\C{13}]$=1.1\%$). The NV centres were created by implantation of N$^+$ ions with an energy of 3.5 keV followed by annealing, resulting in NV centres at a mean depth of 10 nm \cite{Lehtinen2016}. Fig. \ref{Fig:GSLACcomp}a shows a scan across the GSLAC for a particular \NV{14} centre. A reduction in the PL is observed at a field $B_z\approx1024$ G, corresponding to the crossing discussed before. However, full time traces (shown in Fig. \ref{Fig:GSLACcomp}b) now reveal a simple offset of the PL at the crossing, with no obvious oscillatory behaviour. This can be understood by the large magnetic noise originating from the surface, which results in a decay time $T_\times$ shorter than the minimum probe time of $t=1~\mu$s (limited by the lifetime of the singlet state \cite{Manson2006}). The width of the feature in Fig. \ref{Fig:GSLACcomp}a is $\approx1$ G (or $\approx3$ MHz). By measuring various shallow NV centres in the same sample, we found a range of widths of the 1024 G feature from 1 to 3 G (or 3 to 9 MHz). This variability is attributed to different local environments, especially because each NV centre sits at a different distance from the surface. For applications to $T_1$-based NMR spectroscopy as proposed in Ref. \cite{Wood2016}, this implies that nuclear spin species with large gyromagnetic ratios such as $^1$H (Larmor frequency $\approx4.4$ MHz at 1024~G) can be resonantly coupled to a shallow \NV{14} such as that measured in Fig. \ref{Fig:GSLACcomp}a. However, species with smaller gyromagnetic ratios such as \C{13} (Larmor frequency $\approx1.1$~MHz) are generally within the width of the crossing feature in the present sample, and could therefore hardly be detected via cross-relaxation. This motivates further progress in optimising the coherence properties of shallow NV spins, or devising ways to mitigate the effect of dephasing in $T_1$ measurements.

Finally, we measured the properties of \NV{14} centres at the GSLAC in two other settings: (1) deep \NV{14} centres in a CVD diamond with [\C{13}]$=1.1\%$, where decoherence is dominated by the \C{13} bath rather than surface effects; (ii) deep \NV{14} centres in type-Ib diamond grown by the high-pressure high-temperature method, where the main source of decoherence is the bath of electronic spins associated with nitrogen impurities \cite{Hanson2008}. Example scans across the GSLAC are shown in Fig. \ref{Fig:GSLACcomp}c. Deep NVs in CVD diamond showed line widths of the 1024 G feature smaller than $\approx0.3$ G for most NVs ($\approx1$ MHz). By contrast, line widths in the type-Ib diamond are of the order of 10-20 MHz, which makes such diamonds unsuited to $T_1$-based NMR spectroscopy.  

\subsection{\NV{15} centres}
   
\begin{figure}
\includegraphics[width=0.47\textwidth]{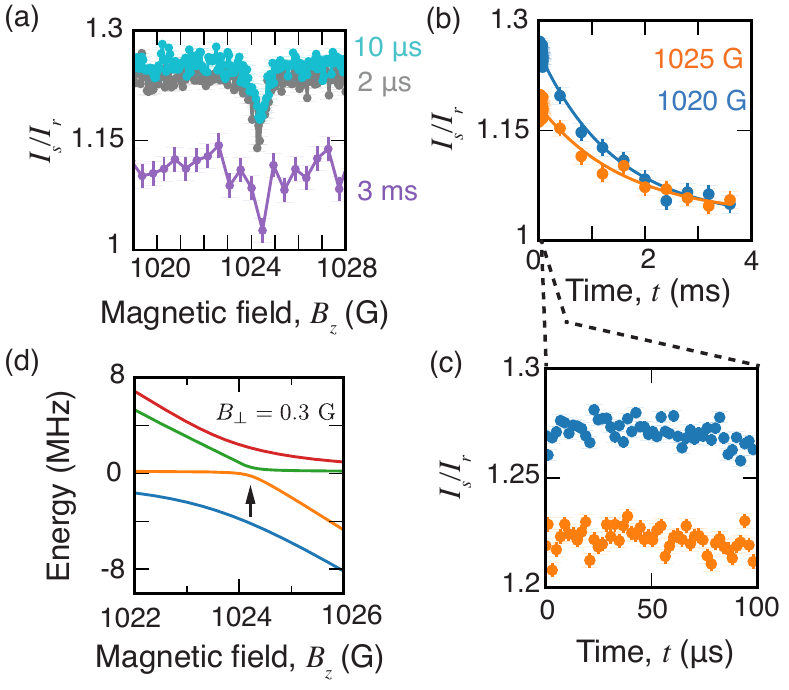}
\caption{(a) PL scan across the GSLAC for a shallow \NV{15} centre in a CVD diamond, with wait times $t=2~\mu$s, $10~\mu$s and 3 ms. A feature at $B_z\approx1024$ G is observed, attributed to spin mixing induced by transverse magnetic fields. (b,c) Time traces recorded at the crossing feature ($B_z\approx1024$ G, orange dots) and away from it ($B_z\approx1020$ G, blue dots). The long and short time scales are shown in (b) and (c), respectively. (d) Energy level structure of the \NV{15} centre in the presence of a transverse magnetic field $B_\perp=0.3$ G, showing an induced avoided crossing (indicated by the arrow).}
\label{Fig:N15dynamics}
\end{figure}
	
As previously discussed, the energy structure of the \NV{15} centre at the GSLAC precludes it from accessing transition frequencies below about 2 MHz. Although this limits the range of nuclear spin species that could be resonantly coupled to the \NV{15}, the highly relevant $^1$H remains accessible. It is therefore important to test the ability to measure the $T_1$ of \NV{15} centres near the GSLAC. 

As in the \NV{14} case, we recorded PL scans across the GSLAC with different wait times, $t$. For this study the measurements were performed on shallow \NV{15} centres in a CVD diamond only, as this is the most relevant sample for sensing applications. The implantation energy was 3.5 keV and the concentration of \C{13} is 1.1\%, similar to the diamond used in Fig. \ref{Fig:GSLACcomp}a. Fig. \ref{Fig:N15dynamics}a shows a scan obtained for a particular \NV{15} centre. The spin population remains essentially constant across the GSLAC, except at a magnetic field $B_z\approx1024$ G where a sharp change is observed. Time traces at and away from the feature are shown in Figs. \ref{Fig:N15dynamics}b and \ref{Fig:N15dynamics}c. While the long time scale reveals an exponential decay with a characteristic time $T_1\approx2$ ms independent from the magnetic field (Fig. \ref{Fig:N15dynamics}b), the contrast of the decay is significantly smaller at $B_z\approx1024$ G. This is due to the initial population being lower, as can be seen from the drop of signal at short time scales (Fig. \ref{Fig:N15dynamics}c). This 1024 G feature was consistently seen in most NVs investigated, exhibiting a variety of amplitudes and widths. At this field, the dominant NV transition has a frequency of $\approx4.3$ MHz. This is 1 G beyond the 1023 G crossing observed in the ODMR (see Fig. \ref{Fig:ODMR}), when the NV transition frequency is $\omega'_\times\approx2.65$ MHz.  

To understand this 1024 G feature, we consider the energy level structure shown in Fig. \ref{Fig:intro}d. As mentioned before, under optical pumping near the GSLAC the \NV{15} centre is efficiently polarised in the state $|0,+1/2\rangle$. This state crosses the state $|-1,-1/2\rangle$ precisely at 1024 G. These two states cannot be coupled directly by a transverse magnetic field because they have distinct nuclear spin projections. However, they are indirectly coupled to each other via transverse-field-enabled coupling to the other two hyperfine-mixed states, which are superpositions of $|0,-1/2\rangle$ and $|-1,+1/2\rangle$. This is illustrated in Fig. \ref{Fig:N15dynamics}d, which shows the computed energy levels as a function of $B_z$ in the presence of a finite transverse field, here $B_\perp=0.3$ G. The transverse field opens a gap between $|0,+1/2\rangle$ and $|-1,-1/2\rangle$ at 1024 G. As a consequence, they become mixed states which can give rise to coherent spin oscillations since optical pumping initialises the NV in the $|0,+1/2\rangle$ state. This situation is reminiscent of the \NV{14} case, where the 1024 G feature was due to an avoided crossing between two states coupled via a transverse magnetic field. The main difference here is that the coupling is indirect, mediated by two intermediate states. In the presence of magnetic noise, the coherent oscillations between the two coupled states are expected to be averaged out and appear as a decrease of the initial spin population, as we observed experimentally in this sample (Fig. \ref{Fig:N15dynamics}c). We note that a transverse-field-induced coupling also occurs at $B_z\approx1027$ G, between $|0,+1/2\rangle$ and $\ket{\beta_+}\approx|0,-1/2\rangle$ (see Fig. \ref{Fig:intro}d). This coupling explains why the dynamic nuclear spin polarisation becomes ineffective around this field, as discussed in section \ref{Sec:ODMR} (see Fig. \ref{Fig:ODMR}).  

An unfortunate consequence of the 1024 G feature of the \NV{15} centre is that resonant coupling with a $^1$H spin would normally occur very close to 1024 G, since the Larmor frequency of $^1$H is $\approx4.36$ MHz at this field. Therefore, any signature of \NV{15}-$^1$H coupling would be overwhelmed by this strong intrinsic feature. It should be noted however that cross-relaxation resonances with nuclear spins should occur on both sides of the GSLAC \cite{Wood2016}, so that $^1$H can still be detected before the GSLAC, at a magnetic field $B_z\approx1022$ G. Moreover, improving the coherence properties (i.e., reducing the noise) of shallow NV centres should significantly reduce the width and amplitude of the 1024 G feature.

\section{All-optical magnetic noise spectroscopy}

\begin{figure}
\includegraphics[width=0.47\textwidth]{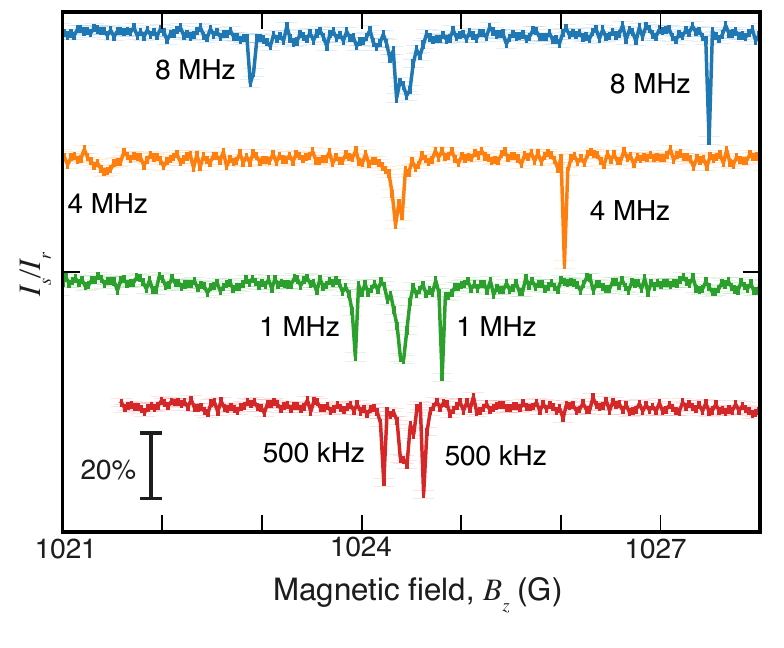}
\caption{PL scans across the GSLAC recorded with a wait time $t=10~\mu$s on the same \NV{14} centre as in Fig. \ref{Fig:C12}. For each scan, a magnetic noise was generated at a central frequency of 8 MHz, 4 MHz, 1 MHz and 500 kHz, respectively (from top down). The root-mean-square amplitude of the applied field is 1 $\mu$T. The curves are offset from each other for clarity.}
\label{Fig: noise}
\end{figure}

By scanning the magnetic field across the GSLAC, we have shown that the transition frequency of the \N{14} centre can be tuned down to frequencies as low as 100 kHz in diamond samples with low intrinsic magnetic noise. This is approximately an order of magnitude below the transition frequencies exhibited by nuclear species (e.g., \C{13}) at 1024 G, thus opening the possibility to perform all-optical NMR spectroscopy by detecting cross-relaxation events between a probe NV spin and target nuclear spins \cite{Wood2016}. When the NV transition frequency is matched to the nuclear Larmor frequency, the fluctuating nuclear field would cause the NV spin to relax faster, translating into a decreased longitudinal relaxation time, $T_1$. 

In order to test the possibility of detecting fluctuating magnetic fields near the GSLAC, we generated a local magnetic field by running an oscillating current through a wire placed in proximity to the diamond. To mimic nuclear spin detection, we applied signals at various frequencies: 8 MHz, 4 MHz ($\sim$ $^1$H or $^{19}$F), 1 MHz ($\sim$ \C{13}), and 500 kHz ($\sim$ $^2$H or $^{17}$O). The alternating current was modulated in amplitude and phase to ensure that the NV is not coherently driven but experiences noise from a randomly fluctuating current around a given frequency, similar to the signal from a possible target nuclear spin. The amplitude of the current was adjusted to obtain a root-mean-square field strength of 1 $\mu$T,  which corresponds approximately to the field generated by a dense organic sample of nuclear spins located at a 5 nm stand-off distance \cite{Staudacher2013}. The probe time was set to $t=10~\mu$s to maximise the PL contrast. The resulting spectra (PL as a function of $B_z$), measured on a deep \N{14} centre in an isotopically-purified CVD diamond, are shown in Fig. \ref{Fig: noise}b. While the 8 MHz, 1 MHz and 500 kHz detections are clear on both sides of the 1024 G crossing, the 4 MHz detection is weaker before the GSLAC. This is due to the NV transition being very weak in this region because of spin mixing (associated with an avoided crossing), as discussed previously (see Fig. \ref{Fig:ODMR}). Past the GSLAC, however, there is no issue measuring any of the signals and thus, NMR spectroscopy would be possible in this region for most commonly found nuclear spin species. We note that the width of the resonances is governed here by the amplitude of the applied field (1 $\mu$T) through power broadening. Weaker signals will produce narrower lines, down to the limit of spectral resolution imposed by spin dephasing, characterised by $T_2^*$ \cite{Hall2016}. Reaching this limit was not possible in our experiment due to limited precision and stability of the applied magnetic field. This could be improved by using, e.g., an electromagnet \cite{Wickenbrock2016}.

\section{Conclusions}

In this work we have investigated the photo-induced spin dynamics of NV centres near the GSLAC. For the \NV{14} centre, the spin transition frequency can be tuned down to values as low as 100 kHz in high purity diamond. At the crossing (1024 G), we observe coherent spin oscillations caused by spin mixing induced by residual transverse magnetic fields. This, in turn, limits the minimum accessible transition frequency exhibited by the environment. Measurements with shallow \NV{14} centres showed that frequencies compatible with nuclear spin signals (1-5 MHz) are within reach. For the \NV{15} centre, the minimum transition frequency practically accessible is of order 2 MHz, governed by the avoided crossing intrinsic to the \NV{15} hyperfine structure. The \NV{15} also exhibits a crossing feature at 1024 G, which is induced by transverse magnetic fields via an indirect hyperfine-mediated process. With this detailed understanding of the low frequency spin dynamics around the GSLAC, we have demonstrated all-optical spectroscopy of externally-generated magnetic noise with frequencies ranging from 8 MHz down to 500 kHz, mimicking signals produced by precessing nuclear spins. This work thus paves the way towards all-optical, nanoscale NMR spectroscopy via cross-relaxation.

\section*{Acknowledgements}

We thank L. McGuinness for experimental assistance with the diamond samples. This work was supported in part by the Australian Research Council (ARC) under the Centre of Excellence scheme (project No. CE110001027). L.C.L.H. acknowledges the support of an ARC Laureate Fellowship (project No. FL130100119).

\bibliographystyle{apsrev4-1}
\bibliography{bib}

\begin{thebibliography}{51}%
\makeatletter
\providecommand \@ifxundefined [1]{%
 \@ifx{#1\undefined}
}%
\providecommand \@ifnum [1]{%
 \ifnum #1\expandafter \@firstoftwo
 \else \expandafter \@secondoftwo
 \fi
}%
\providecommand \@ifx [1]{%
 \ifx #1\expandafter \@firstoftwo
 \else \expandafter \@secondoftwo
 \fi
}%
\providecommand \natexlab [1]{#1}%
\providecommand \enquote  [1]{``#1''}%
\providecommand \bibnamefont  [1]{#1}%
\providecommand \bibfnamefont [1]{#1}%
\providecommand \citenamefont [1]{#1}%
\providecommand \href@noop [0]{\@secondoftwo}%
\providecommand \href [0]{\begingroup \@sanitize@url \@href}%
\providecommand \@href[1]{\@@startlink{#1}\@@href}%
\providecommand \@@href[1]{\endgroup#1\@@endlink}%
\providecommand \@sanitize@url [0]{\catcode `\\12\catcode `\$12\catcode
  `\&12\catcode `\#12\catcode `\^12\catcode `\_12\catcode `\%12\relax}%
\providecommand \@@startlink[1]{}%
\providecommand \@@endlink[0]{}%
\providecommand \url  [0]{\begingroup\@sanitize@url \@url }%
\providecommand \@url [1]{\endgroup\@href {#1}{\urlprefix }}%
\providecommand \urlprefix  [0]{URL }%
\providecommand \Eprint [0]{\href }%
\providecommand \doibase [0]{http://dx.doi.org/}%
\providecommand \selectlanguage [0]{\@gobble}%
\providecommand \bibinfo  [0]{\@secondoftwo}%
\providecommand \bibfield  [0]{\@secondoftwo}%
\providecommand \translation [1]{[#1]}%
\providecommand \BibitemOpen [0]{}%
\providecommand \bibitemStop [0]{}%
\providecommand \bibitemNoStop [0]{.\EOS\space}%
\providecommand \EOS [0]{\spacefactor3000\relax}%
\providecommand \BibitemShut  [1]{\csname bibitem#1\endcsname}%
\let\auto@bib@innerbib\@empty
\bibitem [{\citenamefont {Blank}\ \emph {et~al.}(2003)\citenamefont {Blank},
  \citenamefont {Dunnam}, \citenamefont {Borbat},\ and\ \citenamefont
  {Freed}}]{Blank2003}%
  \BibitemOpen
  \bibfield  {author} {\bibinfo {author} {\bibfnamefont {A.}~\bibnamefont
  {Blank}}, \bibinfo {author} {\bibfnamefont {C.~R.}\ \bibnamefont {Dunnam}},
  \bibinfo {author} {\bibfnamefont {P.~P.}\ \bibnamefont {Borbat}}, \ and\
  \bibinfo {author} {\bibfnamefont {J.~H.}\ \bibnamefont {Freed}},\ }\href
  {\doibase 10.1016/S1090-7807(03)00254-4} {\bibfield  {journal} {\bibinfo
  {journal} {J. Magn. Reson.}\ }\textbf {\bibinfo {volume} {165}},\ \bibinfo
  {pages} {116} (\bibinfo {year} {2003})}\BibitemShut {NoStop}%
\bibitem [{\citenamefont {Poggio}\ and\ \citenamefont
  {Degen}(2010)}]{Poggio2010}%
  \BibitemOpen
  \bibfield  {author} {\bibinfo {author} {\bibfnamefont {M.}~\bibnamefont
  {Poggio}}\ and\ \bibinfo {author} {\bibfnamefont {C.~L.}\ \bibnamefont
  {Degen}},\ }\href {\doibase 10.1088/0957-4484/21/34/342001} {\bibfield
  {journal} {\bibinfo  {journal} {Nanotechnology}\ }\textbf {\bibinfo {volume}
  {21}},\ \bibinfo {pages} {342001} (\bibinfo {year} {2010})}\BibitemShut
  {NoStop}%
\bibitem [{\citenamefont {Artzi}\ \emph {et~al.}(2015)\citenamefont {Artzi},
  \citenamefont {Twig},\ and\ \citenamefont {Blank}}]{Artzi2015}%
  \BibitemOpen
  \bibfield  {author} {\bibinfo {author} {\bibfnamefont {Y.}~\bibnamefont
  {Artzi}}, \bibinfo {author} {\bibfnamefont {Y.}~\bibnamefont {Twig}}, \ and\
  \bibinfo {author} {\bibfnamefont {A.}~\bibnamefont {Blank}},\ }\href
  {\doibase 10.1063/1.4913806} {\bibfield  {journal} {\bibinfo  {journal}
  {Appl. Phys. Lett.}\ }\textbf {\bibinfo {volume} {106}},\ \bibinfo {pages}
  {084104} (\bibinfo {year} {2015})}\BibitemShut {NoStop}%
\bibitem [{\citenamefont {Bienfait}\ \emph {et~al.}(2015)\citenamefont
  {Bienfait}, \citenamefont {Pla}, \citenamefont {Kubo}, \citenamefont {Stern},
  \citenamefont {Zhou}, \citenamefont {Lo}, \citenamefont {Weis}, \citenamefont
  {Schenkel}, \citenamefont {Thewalt}, \citenamefont {Vion}, \citenamefont
  {Esteve}, \citenamefont {Julsgaard}, \citenamefont {M{\o}lmer}, \citenamefont
  {Morton},\ and\ \citenamefont {Bertet}}]{Bienfait2016}%
  \BibitemOpen
  \bibfield  {author} {\bibinfo {author} {\bibfnamefont {A.}~\bibnamefont
  {Bienfait}}, \bibinfo {author} {\bibfnamefont {J.~J.}\ \bibnamefont {Pla}},
  \bibinfo {author} {\bibfnamefont {Y.}~\bibnamefont {Kubo}}, \bibinfo {author}
  {\bibfnamefont {M.}~\bibnamefont {Stern}}, \bibinfo {author} {\bibfnamefont
  {X.}~\bibnamefont {Zhou}}, \bibinfo {author} {\bibfnamefont {C.~C.}\
  \bibnamefont {Lo}}, \bibinfo {author} {\bibfnamefont {C.~D.}\ \bibnamefont
  {Weis}}, \bibinfo {author} {\bibfnamefont {T.}~\bibnamefont {Schenkel}},
  \bibinfo {author} {\bibfnamefont {M.~L.~W.}\ \bibnamefont {Thewalt}},
  \bibinfo {author} {\bibfnamefont {D.}~\bibnamefont {Vion}}, \bibinfo {author}
  {\bibfnamefont {D.}~\bibnamefont {Esteve}}, \bibinfo {author} {\bibfnamefont
  {B.}~\bibnamefont {Julsgaard}}, \bibinfo {author} {\bibfnamefont
  {K.}~\bibnamefont {M{\o}lmer}}, \bibinfo {author} {\bibfnamefont {J.~J.~L.}\
  \bibnamefont {Morton}}, \ and\ \bibinfo {author} {\bibfnamefont
  {P.}~\bibnamefont {Bertet}},\ }\href {\doibase 10.1038/nnano.2015.282}
  {\bibfield  {journal} {\bibinfo  {journal} {Nature Nanotechnology}\ }\textbf
  {\bibinfo {volume} {11}},\ \bibinfo {pages} {253} (\bibinfo {year}
  {2015})}\BibitemShut {NoStop}%
\bibitem [{\citenamefont {Rondin}\ \emph {et~al.}(2014)\citenamefont {Rondin},
  \citenamefont {Tetienne}, \citenamefont {Hingant}, \citenamefont {Roch},
  \citenamefont {Maletinsky},\ and\ \citenamefont {Jacques}}]{Rondin2014}%
  \BibitemOpen
  \bibfield  {author} {\bibinfo {author} {\bibfnamefont {L.}~\bibnamefont
  {Rondin}}, \bibinfo {author} {\bibfnamefont {J.-P.}\ \bibnamefont
  {Tetienne}}, \bibinfo {author} {\bibfnamefont {T.}~\bibnamefont {Hingant}},
  \bibinfo {author} {\bibfnamefont {J.-F.}\ \bibnamefont {Roch}}, \bibinfo
  {author} {\bibfnamefont {P.}~\bibnamefont {Maletinsky}}, \ and\ \bibinfo
  {author} {\bibfnamefont {V.}~\bibnamefont {Jacques}},\ }\href {\doibase
  10.1088/0034-4885/77/5/056503} {\bibfield  {journal} {\bibinfo  {journal}
  {Reports on Progress in Physics}\ }\textbf {\bibinfo {volume} {77}},\
  \bibinfo {pages} {056503} (\bibinfo {year} {2014})}\BibitemShut {NoStop}%
\bibitem [{\citenamefont {Schirhagl}\ \emph {et~al.}(2014)\citenamefont
  {Schirhagl}, \citenamefont {Chang}, \citenamefont {Loretz},\ and\
  \citenamefont {Degen}}]{Schirhagl2014}%
  \BibitemOpen
  \bibfield  {author} {\bibinfo {author} {\bibfnamefont {R.}~\bibnamefont
  {Schirhagl}}, \bibinfo {author} {\bibfnamefont {K.}~\bibnamefont {Chang}},
  \bibinfo {author} {\bibfnamefont {M.}~\bibnamefont {Loretz}}, \ and\ \bibinfo
  {author} {\bibfnamefont {C.~L.}\ \bibnamefont {Degen}},\ }\href {\doibase
  10.1146/annurev-physchem-040513-103659} {\bibfield  {journal} {\bibinfo
  {journal} {Annual Review of Physical Chemistry}\ }\textbf {\bibinfo {volume}
  {65}},\ \bibinfo {pages} {83} (\bibinfo {year} {2014})}\BibitemShut {NoStop}%
\bibitem [{\citenamefont {Doherty}\ \emph {et~al.}(2013)\citenamefont
  {Doherty}, \citenamefont {Manson}, \citenamefont {Delaney}, \citenamefont
  {Jelezko}, \citenamefont {Wrachtrup},\ and\ \citenamefont
  {Hollenberg}}]{Doherty2013}%
  \BibitemOpen
  \bibfield  {author} {\bibinfo {author} {\bibfnamefont {M.~W.}\ \bibnamefont
  {Doherty}}, \bibinfo {author} {\bibfnamefont {N.~B.}\ \bibnamefont {Manson}},
  \bibinfo {author} {\bibfnamefont {P.}~\bibnamefont {Delaney}}, \bibinfo
  {author} {\bibfnamefont {F.}~\bibnamefont {Jelezko}}, \bibinfo {author}
  {\bibfnamefont {J.}~\bibnamefont {Wrachtrup}}, \ and\ \bibinfo {author}
  {\bibfnamefont {L.~C.~L.}\ \bibnamefont {Hollenberg}},\ }\href {\doibase
  10.1016/j.physrep.2013.02.001} {\bibfield  {journal} {\bibinfo  {journal}
  {Physics Reports}\ }\textbf {\bibinfo {volume} {528}},\ \bibinfo {pages} {1}
  (\bibinfo {year} {2013})}\BibitemShut {NoStop}%
\bibitem [{\citenamefont {McGuinness}\ \emph {et~al.}(2011)\citenamefont
  {McGuinness}, \citenamefont {Yan}, \citenamefont {Stacey}, \citenamefont
  {Simpson}, \citenamefont {Hall}, \citenamefont {Maclaurin}, \citenamefont
  {Prawer}, \citenamefont {Mulvaney}, \citenamefont {Wrachtrup}, \citenamefont
  {Caruso}, \citenamefont {Scholten},\ and\ \citenamefont
  {Hollenberg}}]{McGuinness2011}%
  \BibitemOpen
  \bibfield  {author} {\bibinfo {author} {\bibfnamefont {L.~P.}\ \bibnamefont
  {McGuinness}}, \bibinfo {author} {\bibfnamefont {Y.}~\bibnamefont {Yan}},
  \bibinfo {author} {\bibfnamefont {A.}~\bibnamefont {Stacey}}, \bibinfo
  {author} {\bibfnamefont {D.~A.}\ \bibnamefont {Simpson}}, \bibinfo {author}
  {\bibfnamefont {L.~T.}\ \bibnamefont {Hall}}, \bibinfo {author}
  {\bibfnamefont {D.}~\bibnamefont {Maclaurin}}, \bibinfo {author}
  {\bibfnamefont {S.}~\bibnamefont {Prawer}}, \bibinfo {author} {\bibfnamefont
  {P.}~\bibnamefont {Mulvaney}}, \bibinfo {author} {\bibfnamefont
  {J.}~\bibnamefont {Wrachtrup}}, \bibinfo {author} {\bibfnamefont
  {F.}~\bibnamefont {Caruso}}, \bibinfo {author} {\bibfnamefont {R.~E.}\
  \bibnamefont {Scholten}}, \ and\ \bibinfo {author} {\bibfnamefont {L.~C.~L.}\
  \bibnamefont {Hollenberg}},\ }\href {\doibase 10.1038/nnano.2011.64}
  {\bibfield  {journal} {\bibinfo  {journal} {Nature Nanotechnology}\ }\textbf
  {\bibinfo {volume} {6}},\ \bibinfo {pages} {358} (\bibinfo {year}
  {2011})}\BibitemShut {NoStop}%
\bibitem [{\citenamefont {Steinert}\ \emph {et~al.}(2013)\citenamefont
  {Steinert}, \citenamefont {Ziem}, \citenamefont {Hall}, \citenamefont
  {Zappe}, \citenamefont {Schweikert}, \citenamefont {G{\"{o}}tz},
  \citenamefont {Aird}, \citenamefont {Balasubramanian}, \citenamefont
  {Hollenberg},\ and\ \citenamefont {Wrachtrup}}]{Steinert2013}%
  \BibitemOpen
  \bibfield  {author} {\bibinfo {author} {\bibfnamefont {S.}~\bibnamefont
  {Steinert}}, \bibinfo {author} {\bibfnamefont {F.}~\bibnamefont {Ziem}},
  \bibinfo {author} {\bibfnamefont {L.~T.}\ \bibnamefont {Hall}}, \bibinfo
  {author} {\bibfnamefont {A.}~\bibnamefont {Zappe}}, \bibinfo {author}
  {\bibfnamefont {M.}~\bibnamefont {Schweikert}}, \bibinfo {author}
  {\bibfnamefont {N.}~\bibnamefont {G{\"{o}}tz}}, \bibinfo {author}
  {\bibfnamefont {A.}~\bibnamefont {Aird}}, \bibinfo {author} {\bibfnamefont
  {G.}~\bibnamefont {Balasubramanian}}, \bibinfo {author} {\bibfnamefont
  {L.}~\bibnamefont {Hollenberg}}, \ and\ \bibinfo {author} {\bibfnamefont
  {J.}~\bibnamefont {Wrachtrup}},\ }\href {\doibase 10.1038/ncomms2588}
  {\bibfield  {journal} {\bibinfo  {journal} {Nature Communications}\ }\textbf
  {\bibinfo {volume} {4}},\ \bibinfo {pages} {1607} (\bibinfo {year}
  {2013})}\BibitemShut {NoStop}%
\bibitem [{\citenamefont {Tetienne}\ \emph {et~al.}(2013)\citenamefont
  {Tetienne}, \citenamefont {Hingant}, \citenamefont {Rondin}, \citenamefont
  {Cavaill{\`{e}}s}, \citenamefont {Mayer}, \citenamefont {Dantelle},
  \citenamefont {Gacoin}, \citenamefont {Wrachtrup}, \citenamefont {Roch},\
  and\ \citenamefont {Jacques}}]{Tetienne2013}%
  \BibitemOpen
  \bibfield  {author} {\bibinfo {author} {\bibfnamefont {J.-P.}\ \bibnamefont
  {Tetienne}}, \bibinfo {author} {\bibfnamefont {T.}~\bibnamefont {Hingant}},
  \bibinfo {author} {\bibfnamefont {L.}~\bibnamefont {Rondin}}, \bibinfo
  {author} {\bibfnamefont {A.}~\bibnamefont {Cavaill{\`{e}}s}}, \bibinfo
  {author} {\bibfnamefont {L.}~\bibnamefont {Mayer}}, \bibinfo {author}
  {\bibfnamefont {G.}~\bibnamefont {Dantelle}}, \bibinfo {author}
  {\bibfnamefont {T.}~\bibnamefont {Gacoin}}, \bibinfo {author} {\bibfnamefont
  {J.}~\bibnamefont {Wrachtrup}}, \bibinfo {author} {\bibfnamefont {J.-F.}\
  \bibnamefont {Roch}}, \ and\ \bibinfo {author} {\bibfnamefont
  {V.}~\bibnamefont {Jacques}},\ }\href {\doibase 10.1103/PhysRevB.87.235436}
  {\bibfield  {journal} {\bibinfo  {journal} {Physical Review B}\ }\textbf
  {\bibinfo {volume} {87}},\ \bibinfo {pages} {235436} (\bibinfo {year}
  {2013})}\BibitemShut {NoStop}%
\bibitem [{\citenamefont {Kaufmann}\ \emph {et~al.}(2013)\citenamefont
  {Kaufmann}, \citenamefont {Simpson}, \citenamefont {Hall}, \citenamefont
  {Perunicic}, \citenamefont {Senn}, \citenamefont {Steinert}, \citenamefont
  {McGuinness}, \citenamefont {Johnson}, \citenamefont {Ohshima}, \citenamefont
  {Caruso}, \citenamefont {Wrachtrup}, \citenamefont {Scholten}, \citenamefont
  {Mulvaney},\ and\ \citenamefont {Hollenberg}}]{Kaufmann2013}%
  \BibitemOpen
  \bibfield  {author} {\bibinfo {author} {\bibfnamefont {S.}~\bibnamefont
  {Kaufmann}}, \bibinfo {author} {\bibfnamefont {D.~A.}\ \bibnamefont
  {Simpson}}, \bibinfo {author} {\bibfnamefont {L.~T.}\ \bibnamefont {Hall}},
  \bibinfo {author} {\bibfnamefont {V.}~\bibnamefont {Perunicic}}, \bibinfo
  {author} {\bibfnamefont {P.}~\bibnamefont {Senn}}, \bibinfo {author}
  {\bibfnamefont {S.}~\bibnamefont {Steinert}}, \bibinfo {author}
  {\bibfnamefont {L.~P.}\ \bibnamefont {McGuinness}}, \bibinfo {author}
  {\bibfnamefont {B.~C.}\ \bibnamefont {Johnson}}, \bibinfo {author}
  {\bibfnamefont {T.}~\bibnamefont {Ohshima}}, \bibinfo {author} {\bibfnamefont
  {F.}~\bibnamefont {Caruso}}, \bibinfo {author} {\bibfnamefont
  {J.}~\bibnamefont {Wrachtrup}}, \bibinfo {author} {\bibfnamefont {R.~E.}\
  \bibnamefont {Scholten}}, \bibinfo {author} {\bibfnamefont {P.}~\bibnamefont
  {Mulvaney}}, \ and\ \bibinfo {author} {\bibfnamefont {L.}~\bibnamefont
  {Hollenberg}},\ }\href {\doibase 10.1073/pnas.1300640110} {\bibfield
  {journal} {\bibinfo  {journal} {Proceedings of the National Academy of
  Sciences}\ }\textbf {\bibinfo {volume} {110}},\ \bibinfo {pages} {10894}
  (\bibinfo {year} {2013})}\BibitemShut {NoStop}%
\bibitem [{\citenamefont {Ermakova}\ \emph {et~al.}(2013)\citenamefont
  {Ermakova}, \citenamefont {Pramanik}, \citenamefont {Cai}, \citenamefont
  {Algara-Siller}, \citenamefont {Kaiser}, \citenamefont {Weil}, \citenamefont
  {Tzeng}, \citenamefont {Chang}, \citenamefont {McGuinness}, \citenamefont
  {Plenio}, \citenamefont {Naydenov},\ and\ \citenamefont
  {Jelezko}}]{Ermakova2013}%
  \BibitemOpen
  \bibfield  {author} {\bibinfo {author} {\bibfnamefont {A.}~\bibnamefont
  {Ermakova}}, \bibinfo {author} {\bibfnamefont {G.}~\bibnamefont {Pramanik}},
  \bibinfo {author} {\bibfnamefont {J.-M.}\ \bibnamefont {Cai}}, \bibinfo
  {author} {\bibfnamefont {G.}~\bibnamefont {Algara-Siller}}, \bibinfo {author}
  {\bibfnamefont {U.}~\bibnamefont {Kaiser}}, \bibinfo {author} {\bibfnamefont
  {T.}~\bibnamefont {Weil}}, \bibinfo {author} {\bibfnamefont {Y.-K.}\
  \bibnamefont {Tzeng}}, \bibinfo {author} {\bibfnamefont {H.~C.}\ \bibnamefont
  {Chang}}, \bibinfo {author} {\bibfnamefont {L.~P.}\ \bibnamefont
  {McGuinness}}, \bibinfo {author} {\bibfnamefont {M.~B.}\ \bibnamefont
  {Plenio}}, \bibinfo {author} {\bibfnamefont {B.}~\bibnamefont {Naydenov}}, \
  and\ \bibinfo {author} {\bibfnamefont {F.}~\bibnamefont {Jelezko}},\ }\href
  {\doibase 10.1021/nl4015233} {\bibfield  {journal} {\bibinfo  {journal} {Nano
  Letters}\ }\textbf {\bibinfo {volume} {13}},\ \bibinfo {pages} {3305}
  (\bibinfo {year} {2013})}\BibitemShut {NoStop}%
\bibitem [{\citenamefont {Sushkov}\ \emph {et~al.}(2014)\citenamefont
  {Sushkov}, \citenamefont {Chisholm}, \citenamefont {Lovchinsky},
  \citenamefont {Kubo}, \citenamefont {Lo}, \citenamefont {Bennett},
  \citenamefont {Hunger}, \citenamefont {Akimov}, \citenamefont {Walsworth},
  \citenamefont {Park},\ and\ \citenamefont {Lukin}}]{Sushkov2014}%
  \BibitemOpen
  \bibfield  {author} {\bibinfo {author} {\bibfnamefont {A.~O.}\ \bibnamefont
  {Sushkov}}, \bibinfo {author} {\bibfnamefont {N.}~\bibnamefont {Chisholm}},
  \bibinfo {author} {\bibfnamefont {I.}~\bibnamefont {Lovchinsky}}, \bibinfo
  {author} {\bibfnamefont {M.}~\bibnamefont {Kubo}}, \bibinfo {author}
  {\bibfnamefont {P.~K.}\ \bibnamefont {Lo}}, \bibinfo {author} {\bibfnamefont
  {S.~D.}\ \bibnamefont {Bennett}}, \bibinfo {author} {\bibfnamefont
  {D.}~\bibnamefont {Hunger}}, \bibinfo {author} {\bibfnamefont
  {A.}~\bibnamefont {Akimov}}, \bibinfo {author} {\bibfnamefont {R.~L.}\
  \bibnamefont {Walsworth}}, \bibinfo {author} {\bibfnamefont {H.}~\bibnamefont
  {Park}}, \ and\ \bibinfo {author} {\bibfnamefont {M.~D.}\ \bibnamefont
  {Lukin}},\ }\href {\doibase 10.1021/nl502988n} {\bibfield  {journal}
  {\bibinfo  {journal} {Nano Letters}\ }\textbf {\bibinfo {volume} {14}},\
  \bibinfo {pages} {6443} (\bibinfo {year} {2014})}\BibitemShut {NoStop}%
\bibitem [{\citenamefont {Maze}\ \emph {et~al.}(2008)\citenamefont {Maze},
  \citenamefont {Stanwix}, \citenamefont {Hodges}, \citenamefont {Hong},
  \citenamefont {Taylor}, \citenamefont {Cappellaro}, \citenamefont {Jiang},
  \citenamefont {Dutt}, \citenamefont {Togan}, \citenamefont {Zibrov},
  \citenamefont {Yacoby}, \citenamefont {Walsworth},\ and\ \citenamefont
  {Lukin}}]{Maze2008}%
  \BibitemOpen
  \bibfield  {author} {\bibinfo {author} {\bibfnamefont {J.~R.}\ \bibnamefont
  {Maze}}, \bibinfo {author} {\bibfnamefont {P.~L.}\ \bibnamefont {Stanwix}},
  \bibinfo {author} {\bibfnamefont {J.~S.}\ \bibnamefont {Hodges}}, \bibinfo
  {author} {\bibfnamefont {S.}~\bibnamefont {Hong}}, \bibinfo {author}
  {\bibfnamefont {J.~M.}\ \bibnamefont {Taylor}}, \bibinfo {author}
  {\bibfnamefont {P.}~\bibnamefont {Cappellaro}}, \bibinfo {author}
  {\bibfnamefont {L.}~\bibnamefont {Jiang}}, \bibinfo {author} {\bibfnamefont
  {M.~V.~G.}\ \bibnamefont {Dutt}}, \bibinfo {author} {\bibfnamefont
  {E.}~\bibnamefont {Togan}}, \bibinfo {author} {\bibfnamefont {A.~S.}\
  \bibnamefont {Zibrov}}, \bibinfo {author} {\bibfnamefont {A.}~\bibnamefont
  {Yacoby}}, \bibinfo {author} {\bibfnamefont {R.~L.}\ \bibnamefont
  {Walsworth}}, \ and\ \bibinfo {author} {\bibfnamefont {M.~D.}\ \bibnamefont
  {Lukin}},\ }\href {\doibase 10.1038/nature07279} {\bibfield  {journal}
  {\bibinfo  {journal} {Nature}\ }\textbf {\bibinfo {volume} {455}},\ \bibinfo
  {pages} {644} (\bibinfo {year} {2008})}\BibitemShut {NoStop}%
\bibitem [{\citenamefont {{De Lange}}\ \emph {et~al.}(2011)\citenamefont {{De
  Lange}}, \citenamefont {Rist{\`{e}}}, \citenamefont {Dobrovitski},\ and\
  \citenamefont {Hanson}}]{DeLange2011}%
  \BibitemOpen
  \bibfield  {author} {\bibinfo {author} {\bibfnamefont {G.}~\bibnamefont {{De
  Lange}}}, \bibinfo {author} {\bibfnamefont {D.}~\bibnamefont {Rist{\`{e}}}},
  \bibinfo {author} {\bibfnamefont {V.~V.}\ \bibnamefont {Dobrovitski}}, \ and\
  \bibinfo {author} {\bibfnamefont {R.}~\bibnamefont {Hanson}},\ }\href
  {\doibase 10.1103/PhysRevLett.106.080802} {\bibfield  {journal} {\bibinfo
  {journal} {Phys. Rev. Lett.}\ }\textbf {\bibinfo {volume} {106}},\ \bibinfo
  {pages} {1} (\bibinfo {year} {2011})},\ \Eprint
  {http://arxiv.org/abs/1008.4395} {1008.4395} \BibitemShut {NoStop}%
\bibitem [{\citenamefont {McGuinness}\ \emph {et~al.}(2013)\citenamefont
  {McGuinness}, \citenamefont {Hall}, \citenamefont {Stacey}, \citenamefont
  {Simpson}, \citenamefont {Hill}, \citenamefont {Cole}, \citenamefont
  {Ganesan}, \citenamefont {Gibson}, \citenamefont {Prawer}, \citenamefont
  {Mulvaney}, \citenamefont {Jelezko}, \citenamefont {Wrachtrup}, \citenamefont
  {Scholten},\ and\ \citenamefont {Hollenberg}}]{McGuinness2013}%
  \BibitemOpen
  \bibfield  {author} {\bibinfo {author} {\bibfnamefont {L.~P.}\ \bibnamefont
  {McGuinness}}, \bibinfo {author} {\bibfnamefont {L.~T.}\ \bibnamefont
  {Hall}}, \bibinfo {author} {\bibfnamefont {a.}~\bibnamefont {Stacey}},
  \bibinfo {author} {\bibfnamefont {D.~a.}\ \bibnamefont {Simpson}}, \bibinfo
  {author} {\bibfnamefont {C.~D.}\ \bibnamefont {Hill}}, \bibinfo {author}
  {\bibfnamefont {J.~H.}\ \bibnamefont {Cole}}, \bibinfo {author}
  {\bibfnamefont {K.}~\bibnamefont {Ganesan}}, \bibinfo {author} {\bibfnamefont
  {B.~C.}\ \bibnamefont {Gibson}}, \bibinfo {author} {\bibfnamefont
  {S.}~\bibnamefont {Prawer}}, \bibinfo {author} {\bibfnamefont
  {P.}~\bibnamefont {Mulvaney}}, \bibinfo {author} {\bibfnamefont
  {F.}~\bibnamefont {Jelezko}}, \bibinfo {author} {\bibfnamefont
  {J.}~\bibnamefont {Wrachtrup}}, \bibinfo {author} {\bibfnamefont {R.~E.}\
  \bibnamefont {Scholten}}, \ and\ \bibinfo {author} {\bibfnamefont {L.~C.~L.}\
  \bibnamefont {Hollenberg}},\ }\href@noop {} {\bibfield  {journal} {\bibinfo
  {journal} {New J. Phys.}\ }\textbf {\bibinfo {volume} {15}} (\bibinfo {year}
  {2013})},\ \Eprint {http://arxiv.org/abs/1211.5749} {1211.5749} \BibitemShut
  {NoStop}%
\bibitem [{\citenamefont {Cole}\ and\ \citenamefont
  {Hollenberg}(2009)}]{Cole2009}%
  \BibitemOpen
  \bibfield  {author} {\bibinfo {author} {\bibfnamefont {J.~H.}\ \bibnamefont
  {Cole}}\ and\ \bibinfo {author} {\bibfnamefont {L.~C.~L.}\ \bibnamefont
  {Hollenberg}},\ }\href {\doibase 10.1088/0957-4484/20/49/495401} {\bibfield
  {journal} {\bibinfo  {journal} {Nanotechnology}\ }\textbf {\bibinfo {volume}
  {20}},\ \bibinfo {pages} {495401} (\bibinfo {year} {2009})}\BibitemShut
  {NoStop}%
\bibitem [{\citenamefont {Hall}\ \emph {et~al.}(2009)\citenamefont {Hall},
  \citenamefont {Cole}, \citenamefont {Hill},\ and\ \citenamefont
  {Hollenberg}}]{Hall2009}%
  \BibitemOpen
  \bibfield  {author} {\bibinfo {author} {\bibfnamefont {L.~T.}\ \bibnamefont
  {Hall}}, \bibinfo {author} {\bibfnamefont {J.~H.}\ \bibnamefont {Cole}},
  \bibinfo {author} {\bibfnamefont {C.~D.}\ \bibnamefont {Hill}}, \ and\
  \bibinfo {author} {\bibfnamefont {L.~C.~L.}\ \bibnamefont {Hollenberg}},\
  }\href {\doibase 10.1103/PhysRevLett.103.220802} {\bibfield  {journal}
  {\bibinfo  {journal} {Physical Review Letters}\ }\textbf {\bibinfo {volume}
  {103}},\ \bibinfo {pages} {220802} (\bibinfo {year} {2009})}\BibitemShut
  {NoStop}%
\bibitem [{\citenamefont {Laraoui}\ \emph {et~al.}(2010)\citenamefont
  {Laraoui}, \citenamefont {Hodges},\ and\ \citenamefont
  {Meriles}}]{Laraoui2010}%
  \BibitemOpen
  \bibfield  {author} {\bibinfo {author} {\bibfnamefont {A.}~\bibnamefont
  {Laraoui}}, \bibinfo {author} {\bibfnamefont {J.~S.}\ \bibnamefont {Hodges}},
  \ and\ \bibinfo {author} {\bibfnamefont {C.~a.}\ \bibnamefont {Meriles}},\
  }\href {\doibase 10.1063/1.3497004} {\bibfield  {journal} {\bibinfo
  {journal} {Appl. Phys. Lett.}\ }\textbf {\bibinfo {volume} {97}},\ \bibinfo
  {pages} {30} (\bibinfo {year} {2010})},\ \Eprint
  {http://arxiv.org/abs/1009.0316} {1009.0316} \BibitemShut {NoStop}%
\bibitem [{\citenamefont {Grotz}\ \emph {et~al.}(2011)\citenamefont {Grotz},
  \citenamefont {Beck}, \citenamefont {Neumann}, \citenamefont {Naydenov},
  \citenamefont {Reuter}, \citenamefont {Reinhard}, \citenamefont {Jelezko},
  \citenamefont {Wrachtrup}, \citenamefont {Schweinfurth}, \citenamefont
  {Sarkar},\ and\ \citenamefont {Hemmer}}]{Grotz2011}%
  \BibitemOpen
  \bibfield  {author} {\bibinfo {author} {\bibfnamefont {B.}~\bibnamefont
  {Grotz}}, \bibinfo {author} {\bibfnamefont {J.}~\bibnamefont {Beck}},
  \bibinfo {author} {\bibfnamefont {P.}~\bibnamefont {Neumann}}, \bibinfo
  {author} {\bibfnamefont {B.}~\bibnamefont {Naydenov}}, \bibinfo {author}
  {\bibfnamefont {R.}~\bibnamefont {Reuter}}, \bibinfo {author} {\bibfnamefont
  {F.}~\bibnamefont {Reinhard}}, \bibinfo {author} {\bibfnamefont
  {F.}~\bibnamefont {Jelezko}}, \bibinfo {author} {\bibfnamefont
  {J.}~\bibnamefont {Wrachtrup}}, \bibinfo {author} {\bibfnamefont
  {D.}~\bibnamefont {Schweinfurth}}, \bibinfo {author} {\bibfnamefont
  {B.}~\bibnamefont {Sarkar}}, \ and\ \bibinfo {author} {\bibfnamefont
  {P.}~\bibnamefont {Hemmer}},\ }\href
  {http://iopscience.iop.org.ezp.lib.unimelb.edu.au/1367-2630/13/5/055004
  http://iopscience.iop.org.ezp.lib.unimelb.edu.au/1367-2630/13/5/055004/
  http://iopscience.iop.org.ezp.lib.unimelb.edu.au/1367-2630/13/5/055004/pdf/1367-2630{\_}13{\_}5{\_}055004.pdf}
  {\bibfield  {journal} {\bibinfo  {journal} {New Journal of Physics}\ }\textbf
  {\bibinfo {volume} {13}},\ \bibinfo {pages} {055004} (\bibinfo {year}
  {2011})}\BibitemShut {NoStop}%
\bibitem [{\citenamefont {Laraoui}\ \emph {et~al.}(2012)\citenamefont
  {Laraoui}, \citenamefont {Hodges},\ and\ \citenamefont
  {Meriles}}]{Laraoui2012}%
  \BibitemOpen
  \bibfield  {author} {\bibinfo {author} {\bibfnamefont {A.}~\bibnamefont
  {Laraoui}}, \bibinfo {author} {\bibfnamefont {J.~S.}\ \bibnamefont {Hodges}},
  \ and\ \bibinfo {author} {\bibfnamefont {C.~A.}\ \bibnamefont {Meriles}},\
  }\href {\doibase 10.1021/nl300964g} {\bibfield  {journal} {\bibinfo
  {journal} {Nano Letters}\ }\textbf {\bibinfo {volume} {12}},\ \bibinfo
  {pages} {3477} (\bibinfo {year} {2012})}\BibitemShut {NoStop}%
\bibitem [{\citenamefont {Mamin}\ \emph {et~al.}(2012)\citenamefont {Mamin},
  \citenamefont {Sherwood},\ and\ \citenamefont {Rugar}}]{Mamin2012}%
  \BibitemOpen
  \bibfield  {author} {\bibinfo {author} {\bibfnamefont {H.~J.}\ \bibnamefont
  {Mamin}}, \bibinfo {author} {\bibfnamefont {M.~H.}\ \bibnamefont {Sherwood}},
  \ and\ \bibinfo {author} {\bibfnamefont {D.}~\bibnamefont {Rugar}},\ }\href
  {\doibase 10.1103/PhysRevB.86.195422} {\bibfield  {journal} {\bibinfo
  {journal} {Physical Review B}\ }\textbf {\bibinfo {volume} {86}},\ \bibinfo
  {pages} {195422} (\bibinfo {year} {2012})}\BibitemShut {NoStop}%
\bibitem [{\citenamefont {Knowles}\ \emph {et~al.}(2013)\citenamefont
  {Knowles}, \citenamefont {Kara},\ and\ \citenamefont
  {Atat{\"{u}}re}}]{Knowles2013}%
  \BibitemOpen
  \bibfield  {author} {\bibinfo {author} {\bibfnamefont {H.~S.}\ \bibnamefont
  {Knowles}}, \bibinfo {author} {\bibfnamefont {D.~M.}\ \bibnamefont {Kara}}, \
  and\ \bibinfo {author} {\bibfnamefont {M.}~\bibnamefont {Atat{\"{u}}re}},\
  }\href {\doibase 10.1038/nmat3805} {\bibfield  {journal} {\bibinfo  {journal}
  {Nature Materials}\ }\textbf {\bibinfo {volume} {13}},\ \bibinfo {pages} {21}
  (\bibinfo {year} {2013})}\BibitemShut {NoStop}%
\bibitem [{\citenamefont {Mamin}\ \emph {et~al.}(2013)\citenamefont {Mamin},
  \citenamefont {Kim}, \citenamefont {Sherwood}, \citenamefont {Rettner},
  \citenamefont {Ohno}, \citenamefont {Awschalom},\ and\ \citenamefont
  {Rugar}}]{Mamin2013}%
  \BibitemOpen
  \bibfield  {author} {\bibinfo {author} {\bibfnamefont {H.~J.}\ \bibnamefont
  {Mamin}}, \bibinfo {author} {\bibfnamefont {M.}~\bibnamefont {Kim}}, \bibinfo
  {author} {\bibfnamefont {M.~H.}\ \bibnamefont {Sherwood}}, \bibinfo {author}
  {\bibfnamefont {C.~T.}\ \bibnamefont {Rettner}}, \bibinfo {author}
  {\bibfnamefont {K.}~\bibnamefont {Ohno}}, \bibinfo {author} {\bibfnamefont
  {D.~D.}\ \bibnamefont {Awschalom}}, \ and\ \bibinfo {author} {\bibfnamefont
  {D.}~\bibnamefont {Rugar}},\ }\href {\doibase 10.1126/science.1231540}
  {\bibfield  {journal} {\bibinfo  {journal} {Science}\ }\textbf {\bibinfo
  {volume} {339}},\ \bibinfo {pages} {557} (\bibinfo {year}
  {2013})}\BibitemShut {NoStop}%
\bibitem [{\citenamefont {Staudacher}\ \emph {et~al.}(2013)\citenamefont
  {Staudacher}, \citenamefont {Shi}, \citenamefont {Pezzagna}, \citenamefont
  {Meijer}, \citenamefont {Du}, \citenamefont {Meriles}, \citenamefont
  {Reinhard},\ and\ \citenamefont {Wrachtrup}}]{Staudacher2013}%
  \BibitemOpen
  \bibfield  {author} {\bibinfo {author} {\bibfnamefont {T.}~\bibnamefont
  {Staudacher}}, \bibinfo {author} {\bibfnamefont {F.}~\bibnamefont {Shi}},
  \bibinfo {author} {\bibfnamefont {S.}~\bibnamefont {Pezzagna}}, \bibinfo
  {author} {\bibfnamefont {J.}~\bibnamefont {Meijer}}, \bibinfo {author}
  {\bibfnamefont {J.}~\bibnamefont {Du}}, \bibinfo {author} {\bibfnamefont
  {C.~a.}\ \bibnamefont {Meriles}}, \bibinfo {author} {\bibfnamefont
  {F.}~\bibnamefont {Reinhard}}, \ and\ \bibinfo {author} {\bibfnamefont
  {J.}~\bibnamefont {Wrachtrup}},\ }\href {\doibase 10.1126/science.1231675}
  {\bibfield  {journal} {\bibinfo  {journal} {Science (New York, N.Y.)}\
  }\textbf {\bibinfo {volume} {339}},\ \bibinfo {pages} {561} (\bibinfo {year}
  {2013})}\BibitemShut {NoStop}%
\bibitem [{\citenamefont {Loretz}\ \emph {et~al.}(2014)\citenamefont {Loretz},
  \citenamefont {Pezzagna}, \citenamefont {Meijer},\ and\ \citenamefont
  {Degen}}]{Loretz2014}%
  \BibitemOpen
  \bibfield  {author} {\bibinfo {author} {\bibfnamefont {M.}~\bibnamefont
  {Loretz}}, \bibinfo {author} {\bibfnamefont {S.}~\bibnamefont {Pezzagna}},
  \bibinfo {author} {\bibfnamefont {J.}~\bibnamefont {Meijer}}, \ and\ \bibinfo
  {author} {\bibfnamefont {C.~L.}\ \bibnamefont {Degen}},\ }\href {\doibase
  10.1063/1.4862749} {\bibfield  {journal} {\bibinfo  {journal} {Applied
  Physics Letters}\ }\textbf {\bibinfo {volume} {104}},\ \bibinfo {pages}
  {033102} (\bibinfo {year} {2014})}\BibitemShut {NoStop}%
\bibitem [{\citenamefont {M{\"{u}}ller}\ \emph {et~al.}(2014)\citenamefont
  {M{\"{u}}ller}, \citenamefont {Kong}, \citenamefont {Cai}, \citenamefont
  {Melentijevi{\'{c}}}, \citenamefont {Stacey}, \citenamefont {Markham},
  \citenamefont {Twitchen}, \citenamefont {Isoya}, \citenamefont {Pezzagna},
  \citenamefont {Meijer}, \citenamefont {Du}, \citenamefont {Plenio},
  \citenamefont {Naydenov}, \citenamefont {McGuinness},\ and\ \citenamefont
  {Jelezko}}]{Muller2014}%
  \BibitemOpen
  \bibfield  {author} {\bibinfo {author} {\bibfnamefont {C.}~\bibnamefont
  {M{\"{u}}ller}}, \bibinfo {author} {\bibfnamefont {X.}~\bibnamefont {Kong}},
  \bibinfo {author} {\bibfnamefont {J.-M.}\ \bibnamefont {Cai}}, \bibinfo
  {author} {\bibfnamefont {K.}~\bibnamefont {Melentijevi{\'{c}}}}, \bibinfo
  {author} {\bibfnamefont {A.}~\bibnamefont {Stacey}}, \bibinfo {author}
  {\bibfnamefont {M.}~\bibnamefont {Markham}}, \bibinfo {author} {\bibfnamefont
  {D.}~\bibnamefont {Twitchen}}, \bibinfo {author} {\bibfnamefont
  {J.}~\bibnamefont {Isoya}}, \bibinfo {author} {\bibfnamefont
  {S.}~\bibnamefont {Pezzagna}}, \bibinfo {author} {\bibfnamefont
  {J.}~\bibnamefont {Meijer}}, \bibinfo {author} {\bibfnamefont {J.~F.}\
  \bibnamefont {Du}}, \bibinfo {author} {\bibfnamefont {M.~B.}\ \bibnamefont
  {Plenio}}, \bibinfo {author} {\bibfnamefont {B.}~\bibnamefont {Naydenov}},
  \bibinfo {author} {\bibfnamefont {L.~P.}\ \bibnamefont {McGuinness}}, \ and\
  \bibinfo {author} {\bibfnamefont {F.}~\bibnamefont {Jelezko}},\ }\href
  {\doibase 10.1038/ncomms5703} {\bibfield  {journal} {\bibinfo  {journal}
  {Nature communications}\ }\textbf {\bibinfo {volume} {5}},\ \bibinfo {pages}
  {4703} (\bibinfo {year} {2014})}\BibitemShut {NoStop}%
\bibitem [{\citenamefont {DeVience}\ \emph {et~al.}(2015)\citenamefont
  {DeVience}, \citenamefont {Pham}, \citenamefont {Lovchinsky}, \citenamefont
  {Sushkov}, \citenamefont {Bar-Gill}, \citenamefont {Belthangady},
  \citenamefont {Casola}, \citenamefont {Corbett}, \citenamefont {Zhang},
  \citenamefont {Lukin}, \citenamefont {Park}, \citenamefont {Yacoby},\ and\
  \citenamefont {Walsworth}}]{DeVience2015}%
  \BibitemOpen
  \bibfield  {author} {\bibinfo {author} {\bibfnamefont {S.~J.}\ \bibnamefont
  {DeVience}}, \bibinfo {author} {\bibfnamefont {L.~M.}\ \bibnamefont {Pham}},
  \bibinfo {author} {\bibfnamefont {I.}~\bibnamefont {Lovchinsky}}, \bibinfo
  {author} {\bibfnamefont {A.~O.}\ \bibnamefont {Sushkov}}, \bibinfo {author}
  {\bibfnamefont {N.}~\bibnamefont {Bar-Gill}}, \bibinfo {author}
  {\bibfnamefont {C.}~\bibnamefont {Belthangady}}, \bibinfo {author}
  {\bibfnamefont {F.}~\bibnamefont {Casola}}, \bibinfo {author} {\bibfnamefont
  {M.}~\bibnamefont {Corbett}}, \bibinfo {author} {\bibfnamefont
  {H.}~\bibnamefont {Zhang}}, \bibinfo {author} {\bibfnamefont
  {M.}~\bibnamefont {Lukin}}, \bibinfo {author} {\bibfnamefont
  {H.}~\bibnamefont {Park}}, \bibinfo {author} {\bibfnamefont {A.}~\bibnamefont
  {Yacoby}}, \ and\ \bibinfo {author} {\bibfnamefont {R.~L.}\ \bibnamefont
  {Walsworth}},\ }\href {\doibase 10.1038/nnano.2014.313} {\bibfield  {journal}
  {\bibinfo  {journal} {Nature Nanotechnology}\ }\textbf {\bibinfo {volume}
  {10}},\ \bibinfo {pages} {129} (\bibinfo {year} {2015})}\BibitemShut
  {NoStop}%
\bibitem [{\citenamefont {Jarmola}\ \emph {et~al.}(2012)\citenamefont
  {Jarmola}, \citenamefont {Acosta}, \citenamefont {Jensen}, \citenamefont
  {Chemerisov},\ and\ \citenamefont {Budker}}]{Jarmola2012}%
  \BibitemOpen
  \bibfield  {author} {\bibinfo {author} {\bibfnamefont {A.}~\bibnamefont
  {Jarmola}}, \bibinfo {author} {\bibfnamefont {V.~M.}\ \bibnamefont {Acosta}},
  \bibinfo {author} {\bibfnamefont {K.}~\bibnamefont {Jensen}}, \bibinfo
  {author} {\bibfnamefont {S.}~\bibnamefont {Chemerisov}}, \ and\ \bibinfo
  {author} {\bibfnamefont {D.}~\bibnamefont {Budker}},\ }\href {\doibase
  10.1103/PhysRevLett.108.197601} {\bibfield  {journal} {\bibinfo  {journal}
  {Physical Review Letters}\ }\textbf {\bibinfo {volume} {108}},\ \bibinfo
  {pages} {197601} (\bibinfo {year} {2012})}\BibitemShut {NoStop}%
\bibitem [{\citenamefont {Wang}\ \emph {et~al.}(2014)\citenamefont {Wang},
  \citenamefont {Shin}, \citenamefont {Seltzer}, \citenamefont {Avalos},
  \citenamefont {Pines},\ and\ \citenamefont {Bajaj}}]{Wang2014}%
  \BibitemOpen
  \bibfield  {author} {\bibinfo {author} {\bibfnamefont {H.-J.}\ \bibnamefont
  {Wang}}, \bibinfo {author} {\bibfnamefont {C.~S.}\ \bibnamefont {Shin}},
  \bibinfo {author} {\bibfnamefont {S.~J.}\ \bibnamefont {Seltzer}}, \bibinfo
  {author} {\bibfnamefont {C.~E.}\ \bibnamefont {Avalos}}, \bibinfo {author}
  {\bibfnamefont {A.}~\bibnamefont {Pines}}, \ and\ \bibinfo {author}
  {\bibfnamefont {V.~S.}\ \bibnamefont {Bajaj}},\ }\href {\doibase
  10.1038/ncomms5135} {\bibfield  {journal} {\bibinfo  {journal} {Nat.
  Commun.}\ }\textbf {\bibinfo {volume} {5}},\ \bibinfo {pages} {4135}
  (\bibinfo {year} {2014})},\ \Eprint {http://arxiv.org/abs/1312.6313}
  {1312.6313} \BibitemShut {NoStop}%
\bibitem [{\citenamefont {van~der Sar}\ \emph {et~al.}(2015)\citenamefont
  {van~der Sar}, \citenamefont {Casola}, \citenamefont {Walsworth},\ and\
  \citenamefont {Yacoby}}]{VanderSar2015}%
  \BibitemOpen
  \bibfield  {author} {\bibinfo {author} {\bibfnamefont {T.}~\bibnamefont
  {van~der Sar}}, \bibinfo {author} {\bibfnamefont {F.}~\bibnamefont {Casola}},
  \bibinfo {author} {\bibfnamefont {R.}~\bibnamefont {Walsworth}}, \ and\
  \bibinfo {author} {\bibfnamefont {A.}~\bibnamefont {Yacoby}},\ }\href
  {\doibase 10.1038/ncomms8886} {\bibfield  {journal} {\bibinfo  {journal}
  {Nature Communications}\ }\textbf {\bibinfo {volume} {6}},\ \bibinfo {pages}
  {7886} (\bibinfo {year} {2015})}\BibitemShut {NoStop}%
\bibitem [{\citenamefont {Hall}\ \emph {et~al.}(2016)\citenamefont {Hall},
  \citenamefont {Kehayias}, \citenamefont {Simpson}, \citenamefont {Jarmola},
  \citenamefont {Stacey}, \citenamefont {Budker},\ and\ \citenamefont
  {Hollenberg}}]{Hall2016}%
  \BibitemOpen
  \bibfield  {author} {\bibinfo {author} {\bibfnamefont {L.~T.}\ \bibnamefont
  {Hall}}, \bibinfo {author} {\bibfnamefont {P.}~\bibnamefont {Kehayias}},
  \bibinfo {author} {\bibfnamefont {D.~A.}\ \bibnamefont {Simpson}}, \bibinfo
  {author} {\bibfnamefont {A.}~\bibnamefont {Jarmola}}, \bibinfo {author}
  {\bibfnamefont {A.}~\bibnamefont {Stacey}}, \bibinfo {author} {\bibfnamefont
  {D.}~\bibnamefont {Budker}}, \ and\ \bibinfo {author} {\bibfnamefont
  {L.~C.~L.}\ \bibnamefont {Hollenberg}},\ }\href {\doibase
  10.1038/ncomms10211} {\bibfield  {journal} {\bibinfo  {journal} {Nature
  Communications}\ }\textbf {\bibinfo {volume} {7}},\ \bibinfo {pages} {10211}
  (\bibinfo {year} {2016})}\BibitemShut {NoStop}%
\bibitem [{\citenamefont {Wood}\ \emph {et~al.}(2016)\citenamefont {Wood},
  \citenamefont {Broadway}, \citenamefont {Hall}, \citenamefont {Stacey},
  \citenamefont {Simpson}, \citenamefont {Tetienne},\ and\ \citenamefont
  {Hollenberg}}]{Wood2016}%
  \BibitemOpen
  \bibfield  {author} {\bibinfo {author} {\bibfnamefont {J.~D.~A.}\
  \bibnamefont {Wood}}, \bibinfo {author} {\bibfnamefont {D.~A.}\ \bibnamefont
  {Broadway}}, \bibinfo {author} {\bibfnamefont {L.~T.}\ \bibnamefont {Hall}},
  \bibinfo {author} {\bibfnamefont {A.}~\bibnamefont {Stacey}}, \bibinfo
  {author} {\bibfnamefont {D.~A.}\ \bibnamefont {Simpson}}, \bibinfo {author}
  {\bibfnamefont {J.-P.}\ \bibnamefont {Tetienne}}, \ and\ \bibinfo {author}
  {\bibfnamefont {L.~C.~L.}\ \bibnamefont {Hollenberg}},\ }\href
  {http://arxiv.org/abs/1604.00160} {\  (\bibinfo {year} {2016})},\ \Eprint
  {http://arxiv.org/abs/1604.00160} {arXiv:1604.00160} \BibitemShut {NoStop}%
\bibitem [{\citenamefont {He}\ \emph {et~al.}(1993)\citenamefont {He},
  \citenamefont {Manson},\ and\ \citenamefont {Fisk}}]{He1993}%
  \BibitemOpen
  \bibfield  {author} {\bibinfo {author} {\bibfnamefont {X.-F.}\ \bibnamefont
  {He}}, \bibinfo {author} {\bibfnamefont {N.~B.}\ \bibnamefont {Manson}}, \
  and\ \bibinfo {author} {\bibfnamefont {P.~T.~H.}\ \bibnamefont {Fisk}},\
  }\href {\doibase 10.1103/PhysRevB.47.8809} {\bibfield  {journal} {\bibinfo
  {journal} {Phys. Rev. B}\ }\textbf {\bibinfo {volume} {47}},\ \bibinfo
  {pages} {8809} (\bibinfo {year} {1993})}\BibitemShut {NoStop}%
\bibitem [{\citenamefont {Epstein}\ \emph {et~al.}(2005)\citenamefont
  {Epstein}, \citenamefont {Mendoza}, \citenamefont {Kato},\ and\ \citenamefont
  {Awschalom}}]{Epstein2005}%
  \BibitemOpen
  \bibfield  {author} {\bibinfo {author} {\bibfnamefont {R.~J.}\ \bibnamefont
  {Epstein}}, \bibinfo {author} {\bibfnamefont {F.~M.}\ \bibnamefont
  {Mendoza}}, \bibinfo {author} {\bibfnamefont {Y.~K.}\ \bibnamefont {Kato}}, \
  and\ \bibinfo {author} {\bibfnamefont {D.~D.}\ \bibnamefont {Awschalom}},\
  }\href {\doibase 10.1038/nphys141} {\bibfield  {journal} {\bibinfo  {journal}
  {Nature Physics}\ }\textbf {\bibinfo {volume} {1}},\ \bibinfo {pages} {94}
  (\bibinfo {year} {2005})}\BibitemShut {NoStop}%
\bibitem [{\citenamefont {Wei}\ and\ \citenamefont {Manson}(1999)}]{Wei1999}%
  \BibitemOpen
  \bibfield  {author} {\bibinfo {author} {\bibfnamefont {C.}~\bibnamefont
  {Wei}}\ and\ \bibinfo {author} {\bibfnamefont {N.~B.}\ \bibnamefont
  {Manson}},\ }\href {\doibase 10.1103/PhysRevA.60.2540} {\bibfield  {journal}
  {\bibinfo  {journal} {Phys. Rev. A}\ }\textbf {\bibinfo {volume} {60}},\
  \bibinfo {pages} {2540} (\bibinfo {year} {1999})}\BibitemShut {NoStop}%
\bibitem [{\citenamefont {Wilson}\ \emph {et~al.}(2003)\citenamefont {Wilson},
  \citenamefont {Manson},\ and\ \citenamefont {Wei}}]{Wilson2003}%
  \BibitemOpen
  \bibfield  {author} {\bibinfo {author} {\bibfnamefont {E.~A.}\ \bibnamefont
  {Wilson}}, \bibinfo {author} {\bibfnamefont {N.~B.}\ \bibnamefont {Manson}},
  \ and\ \bibinfo {author} {\bibfnamefont {C.}~\bibnamefont {Wei}},\ }\href
  {\doibase 10.1103/PhysRevA.67.023812} {\bibfield  {journal} {\bibinfo
  {journal} {Phys. Rev. A}\ }\textbf {\bibinfo {volume} {67}},\ \bibinfo
  {pages} {023812} (\bibinfo {year} {2003})}\BibitemShut {NoStop}%
\bibitem [{\citenamefont {Fuchs}\ \emph {et~al.}(2011)\citenamefont {Fuchs},
  \citenamefont {Burkard}, \citenamefont {Klimov},\ and\ \citenamefont
  {Awschalom}}]{Fuchs2011}%
  \BibitemOpen
  \bibfield  {author} {\bibinfo {author} {\bibfnamefont {G.~D.}\ \bibnamefont
  {Fuchs}}, \bibinfo {author} {\bibfnamefont {G.}~\bibnamefont {Burkard}},
  \bibinfo {author} {\bibfnamefont {P.~V.}\ \bibnamefont {Klimov}}, \ and\
  \bibinfo {author} {\bibfnamefont {D.~D.}\ \bibnamefont {Awschalom}},\ }\href
  {\doibase 10.1038/nphys2026} {\bibfield  {journal} {\bibinfo  {journal}
  {Nature Physics}\ }\textbf {\bibinfo {volume} {7}},\ \bibinfo {pages} {789}
  (\bibinfo {year} {2011})}\BibitemShut {NoStop}%
\bibitem [{\citenamefont {Wang}\ \emph {et~al.}(2013)\citenamefont {Wang},
  \citenamefont {Shin}, \citenamefont {Avalos}, \citenamefont {Seltzer},
  \citenamefont {Budker}, \citenamefont {Pines},\ and\ \citenamefont
  {Bajaj}}]{Wang2013}%
  \BibitemOpen
  \bibfield  {author} {\bibinfo {author} {\bibfnamefont {H.-J.}\ \bibnamefont
  {Wang}}, \bibinfo {author} {\bibfnamefont {C.~S.}\ \bibnamefont {Shin}},
  \bibinfo {author} {\bibfnamefont {C.~E.}\ \bibnamefont {Avalos}}, \bibinfo
  {author} {\bibfnamefont {S.~J.}\ \bibnamefont {Seltzer}}, \bibinfo {author}
  {\bibfnamefont {D.}~\bibnamefont {Budker}}, \bibinfo {author} {\bibfnamefont
  {A.}~\bibnamefont {Pines}}, \ and\ \bibinfo {author} {\bibfnamefont {V.~S.}\
  \bibnamefont {Bajaj}},\ }\href
  {http://www.nature.com.ezp.lib.unimelb.edu.au/ncomms/2013/130605/ncomms2930/full/ncomms2930.html
  http://www.nature.com.ezp.lib.unimelb.edu.au/ncomms/2013/130605/ncomms2930/pdf/ncomms2930.pdf}
  {\bibfield  {journal} {\bibinfo  {journal} {Nature Communications}\ }\textbf
  {\bibinfo {volume} {4}},\ \bibinfo {pages} {1940} (\bibinfo {year}
  {2013})}\BibitemShut {NoStop}%
\bibitem [{\citenamefont {Wang}\ \emph {et~al.}(2015)\citenamefont {Wang},
  \citenamefont {Liu},\ and\ \citenamefont {Yang}}]{Wang2015a}%
  \BibitemOpen
  \bibfield  {author} {\bibinfo {author} {\bibfnamefont {P.}~\bibnamefont
  {Wang}}, \bibinfo {author} {\bibfnamefont {B.}~\bibnamefont {Liu}}, \ and\
  \bibinfo {author} {\bibfnamefont {W.}~\bibnamefont {Yang}},\ }\href@noop {}
  {\bibfield  {journal} {\bibinfo  {journal} {Scientific Reports}\ }\textbf
  {\bibinfo {volume} {5}},\ \bibinfo {pages} {15847} (\bibinfo {year}
  {2015})}\BibitemShut {NoStop}%
\bibitem [{\citenamefont {Wickenbrock}\ \emph {et~al.}(2016)\citenamefont
  {Wickenbrock}, \citenamefont {Zheng}, \citenamefont {Bougas}, \citenamefont
  {Leefer}, \citenamefont {Afach}, \citenamefont {Jarmola}, \citenamefont
  {Acosta},\ and\ \citenamefont {Budker}}]{Wickenbrock2016}%
  \BibitemOpen
  \bibfield  {author} {\bibinfo {author} {\bibfnamefont {A.}~\bibnamefont
  {Wickenbrock}}, \bibinfo {author} {\bibfnamefont {H.}~\bibnamefont {Zheng}},
  \bibinfo {author} {\bibfnamefont {L.}~\bibnamefont {Bougas}}, \bibinfo
  {author} {\bibfnamefont {N.}~\bibnamefont {Leefer}}, \bibinfo {author}
  {\bibfnamefont {S.}~\bibnamefont {Afach}}, \bibinfo {author} {\bibfnamefont
  {A.}~\bibnamefont {Jarmola}}, \bibinfo {author} {\bibfnamefont {V.~M.}\
  \bibnamefont {Acosta}}, \ and\ \bibinfo {author} {\bibfnamefont
  {D.}~\bibnamefont {Budker}},\ }\href {http://arxiv.org/abs/1606.03070} {\
  (\bibinfo {year} {2016})},\ \Eprint {http://arxiv.org/abs/1606.03070}
  {arXiv:1606.03070} \BibitemShut {NoStop}%
\bibitem [{\citenamefont {Felton}\ \emph {et~al.}(2009)\citenamefont {Felton},
  \citenamefont {Edmonds}, \citenamefont {Newton}, \citenamefont {Martineau},
  \citenamefont {Fisher}, \citenamefont {Twitchen},\ and\ \citenamefont
  {Baker}}]{Felton2009}%
  \BibitemOpen
  \bibfield  {author} {\bibinfo {author} {\bibfnamefont {S.}~\bibnamefont
  {Felton}}, \bibinfo {author} {\bibfnamefont {A.~M.}\ \bibnamefont {Edmonds}},
  \bibinfo {author} {\bibfnamefont {M.~E.}\ \bibnamefont {Newton}}, \bibinfo
  {author} {\bibfnamefont {P.~M.}\ \bibnamefont {Martineau}}, \bibinfo {author}
  {\bibfnamefont {D.}~\bibnamefont {Fisher}}, \bibinfo {author} {\bibfnamefont
  {D.~J.}\ \bibnamefont {Twitchen}}, \ and\ \bibinfo {author} {\bibfnamefont
  {J.~M.}\ \bibnamefont {Baker}},\ }\href {\doibase 10.1103/PhysRevB.79.075203}
  {\bibfield  {journal} {\bibinfo  {journal} {Physical Review B}\ }\textbf
  {\bibinfo {volume} {79}},\ \bibinfo {pages} {075203} (\bibinfo {year}
  {2009})}\BibitemShut {NoStop}%
\bibitem [{\citenamefont {Gruber}\ \emph {et~al.}(1997)\citenamefont {Gruber},
  \citenamefont {Dr{\"{a}}benstedt}, \citenamefont {Tietz}, \citenamefont
  {Fleury}, \citenamefont {Wrachtrup},\ and\ \citenamefont {von
  Borczyskowski}}]{Gruber1997}%
  \BibitemOpen
  \bibfield  {author} {\bibinfo {author} {\bibfnamefont {A.}~\bibnamefont
  {Gruber}}, \bibinfo {author} {\bibfnamefont {A.}~\bibnamefont
  {Dr{\"{a}}benstedt}}, \bibinfo {author} {\bibfnamefont {C.}~\bibnamefont
  {Tietz}}, \bibinfo {author} {\bibfnamefont {L.}~\bibnamefont {Fleury}},
  \bibinfo {author} {\bibfnamefont {J.}~\bibnamefont {Wrachtrup}}, \ and\
  \bibinfo {author} {\bibfnamefont {C.}~\bibnamefont {von Borczyskowski}},\
  }\href {\doibase 10.1126/science.276.5321.2012} {\bibfield  {journal}
  {\bibinfo  {journal} {Science}\ }\textbf {\bibinfo {volume} {276}},\ \bibinfo
  {pages} {2012} (\bibinfo {year} {1997})}\BibitemShut {NoStop}%
\bibitem [{\citenamefont {Manson}\ \emph {et~al.}(2006)\citenamefont {Manson},
  \citenamefont {Harrison},\ and\ \citenamefont {Sellars}}]{Manson2006}%
  \BibitemOpen
  \bibfield  {author} {\bibinfo {author} {\bibfnamefont {N.~B.}\ \bibnamefont
  {Manson}}, \bibinfo {author} {\bibfnamefont {J.~P.}\ \bibnamefont
  {Harrison}}, \ and\ \bibinfo {author} {\bibfnamefont {M.~J.}\ \bibnamefont
  {Sellars}},\ }\href {\doibase 10.1103/PhysRevB.74.104303} {\bibfield
  {journal} {\bibinfo  {journal} {Physical Review B}\ }\textbf {\bibinfo
  {volume} {74}},\ \bibinfo {pages} {104303} (\bibinfo {year}
  {2006})}\BibitemShut {NoStop}%
\bibitem [{\citenamefont {Jacques}\ \emph {et~al.}(2009)\citenamefont
  {Jacques}, \citenamefont {Neumann}, \citenamefont {Beck}, \citenamefont
  {Markham}, \citenamefont {Twitchen}, \citenamefont {Meijer}, \citenamefont
  {Kaiser}, \citenamefont {Balasubramanian}, \citenamefont {Jelezko},\ and\
  \citenamefont {Wrachtrup}}]{Jacques2009}%
  \BibitemOpen
  \bibfield  {author} {\bibinfo {author} {\bibfnamefont {V.}~\bibnamefont
  {Jacques}}, \bibinfo {author} {\bibfnamefont {P.}~\bibnamefont {Neumann}},
  \bibinfo {author} {\bibfnamefont {J.}~\bibnamefont {Beck}}, \bibinfo {author}
  {\bibfnamefont {M.}~\bibnamefont {Markham}}, \bibinfo {author} {\bibfnamefont
  {D.}~\bibnamefont {Twitchen}}, \bibinfo {author} {\bibfnamefont
  {J.}~\bibnamefont {Meijer}}, \bibinfo {author} {\bibfnamefont
  {F.}~\bibnamefont {Kaiser}}, \bibinfo {author} {\bibfnamefont
  {G.}~\bibnamefont {Balasubramanian}}, \bibinfo {author} {\bibfnamefont
  {F.}~\bibnamefont {Jelezko}}, \ and\ \bibinfo {author} {\bibfnamefont
  {J.}~\bibnamefont {Wrachtrup}},\ }\href {\doibase
  10.1103/PhysRevLett.102.057403} {\bibfield  {journal} {\bibinfo  {journal}
  {Physical Review Letters}\ }\textbf {\bibinfo {volume} {102}},\ \bibinfo
  {pages} {057403} (\bibinfo {year} {2009})}\BibitemShut {NoStop}%
\bibitem [{\citenamefont {Iv{\'{a}}dy}\ \emph {et~al.}(2015)\citenamefont
  {Iv{\'{a}}dy}, \citenamefont {Sz{\'{a}}sz}, \citenamefont {Falk},
  \citenamefont {Klimov}, \citenamefont {Christle}, \citenamefont
  {Janz{\'{e}}n}, \citenamefont {Abrikosov}, \citenamefont {Awschalom},\ and\
  \citenamefont {Gali}}]{Ivady2015}%
  \BibitemOpen
  \bibfield  {author} {\bibinfo {author} {\bibfnamefont {V.}~\bibnamefont
  {Iv{\'{a}}dy}}, \bibinfo {author} {\bibfnamefont {K.}~\bibnamefont
  {Sz{\'{a}}sz}}, \bibinfo {author} {\bibfnamefont {A.~L.}\ \bibnamefont
  {Falk}}, \bibinfo {author} {\bibfnamefont {P.~V.}\ \bibnamefont {Klimov}},
  \bibinfo {author} {\bibfnamefont {D.~J.}\ \bibnamefont {Christle}}, \bibinfo
  {author} {\bibfnamefont {E.}~\bibnamefont {Janz{\'{e}}n}}, \bibinfo {author}
  {\bibfnamefont {I.~A.}\ \bibnamefont {Abrikosov}}, \bibinfo {author}
  {\bibfnamefont {D.~D.}\ \bibnamefont {Awschalom}}, \ and\ \bibinfo {author}
  {\bibfnamefont {A.}~\bibnamefont {Gali}},\ }\href {\doibase
  10.1103/PhysRevB.92.115206} {\bibfield  {journal} {\bibinfo  {journal}
  {Physical Review B}\ }\textbf {\bibinfo {volume} {92}},\ \bibinfo {pages}
  {115206} (\bibinfo {year} {2015})}\BibitemShut {NoStop}%
\bibitem [{\citenamefont {Balasubramanian}\ \emph {et~al.}(2009)\citenamefont
  {Balasubramanian}, \citenamefont {Neumann}, \citenamefont {Twitchen},
  \citenamefont {Markham}, \citenamefont {Kolesov}, \citenamefont {Mizuochi},
  \citenamefont {Isoya}, \citenamefont {Achard}, \citenamefont {Beck},
  \citenamefont {Tissler}, \citenamefont {Jacques}, \citenamefont {Hemmer},
  \citenamefont {Jelezko},\ and\ \citenamefont
  {Wrachtrup}}]{Balasubramanian2009}%
  \BibitemOpen
  \bibfield  {author} {\bibinfo {author} {\bibfnamefont {G.}~\bibnamefont
  {Balasubramanian}}, \bibinfo {author} {\bibfnamefont {P.}~\bibnamefont
  {Neumann}}, \bibinfo {author} {\bibfnamefont {D.}~\bibnamefont {Twitchen}},
  \bibinfo {author} {\bibfnamefont {M.}~\bibnamefont {Markham}}, \bibinfo
  {author} {\bibfnamefont {R.}~\bibnamefont {Kolesov}}, \bibinfo {author}
  {\bibfnamefont {N.}~\bibnamefont {Mizuochi}}, \bibinfo {author}
  {\bibfnamefont {J.}~\bibnamefont {Isoya}}, \bibinfo {author} {\bibfnamefont
  {J.}~\bibnamefont {Achard}}, \bibinfo {author} {\bibfnamefont
  {J.}~\bibnamefont {Beck}}, \bibinfo {author} {\bibfnamefont {J.}~\bibnamefont
  {Tissler}}, \bibinfo {author} {\bibfnamefont {V.}~\bibnamefont {Jacques}},
  \bibinfo {author} {\bibfnamefont {P.~R.}\ \bibnamefont {Hemmer}}, \bibinfo
  {author} {\bibfnamefont {F.}~\bibnamefont {Jelezko}}, \ and\ \bibinfo
  {author} {\bibfnamefont {J.}~\bibnamefont {Wrachtrup}},\ }\href {\doibase
  10.1038/nmat2420} {\bibfield  {journal} {\bibinfo  {journal} {Nature
  Materials}\ }\textbf {\bibinfo {volume} {8}},\ \bibinfo {pages} {383}
  (\bibinfo {year} {2009})}\BibitemShut {NoStop}%
\bibitem [{\citenamefont {Hall}\ \emph {et~al.}(2014)\citenamefont {Hall},
  \citenamefont {Cole},\ and\ \citenamefont {Hollenberg}}]{Hall2014}%
  \BibitemOpen
  \bibfield  {author} {\bibinfo {author} {\bibfnamefont {L.~T.}\ \bibnamefont
  {Hall}}, \bibinfo {author} {\bibfnamefont {J.~H.}\ \bibnamefont {Cole}}, \
  and\ \bibinfo {author} {\bibfnamefont {L.~C.~L.}\ \bibnamefont
  {Hollenberg}},\ }\href {http://link.aps.org/doi/10.1103/PhysRevB.90.075201
  http://journals.aps.org.ezp.lib.unimelb.edu.au/prb/abstract/10.1103/PhysRevB.90.075201}
  {\bibfield  {journal} {\bibinfo  {journal} {Physical Review B}\ }\textbf
  {\bibinfo {volume} {90}},\ \bibinfo {pages} {075201} (\bibinfo {year}
  {2014})}\BibitemShut {NoStop}%
\bibitem [{\citenamefont {Maze}\ \emph {et~al.}(2012)\citenamefont {Maze},
  \citenamefont {Dr{\'e}au}, \citenamefont {Waselowski}, \citenamefont
  {Duarte}, \citenamefont {Roch},\ and\ \citenamefont {Jacques}}]{Maze2012}%
  \BibitemOpen
  \bibfield  {author} {\bibinfo {author} {\bibfnamefont {J.~R.}\ \bibnamefont
  {Maze}}, \bibinfo {author} {\bibfnamefont {A.}~\bibnamefont {Dr{\'e}au}},
  \bibinfo {author} {\bibfnamefont {V.}~\bibnamefont {Waselowski}}, \bibinfo
  {author} {\bibfnamefont {H.}~\bibnamefont {Duarte}}, \bibinfo {author}
  {\bibfnamefont {J.-F.}\ \bibnamefont {Roch}}, \ and\ \bibinfo {author}
  {\bibfnamefont {V.}~\bibnamefont {Jacques}},\ }\href
  {http://stacks.iop.org/1367-2630/14/i=10/a=103041} {\bibfield  {journal}
  {\bibinfo  {journal} {New Journal of Physics}\ }\textbf {\bibinfo {volume}
  {14}},\ \bibinfo {pages} {103041} (\bibinfo {year} {2012})}\BibitemShut
  {NoStop}%
\bibitem [{\citenamefont {Lehtinen}\ \emph {et~al.}(2016)\citenamefont
  {Lehtinen}, \citenamefont {Naydenov}, \citenamefont {B{\"{o}}rner},
  \citenamefont {Melentjevic}, \citenamefont {M{\"{u}}ller}, \citenamefont
  {McGuinness}, \citenamefont {Pezzagna}, \citenamefont {Meijer}, \citenamefont
  {Kaiser},\ and\ \citenamefont {Jelezko}}]{Lehtinen2016}%
  \BibitemOpen
  \bibfield  {author} {\bibinfo {author} {\bibfnamefont {O.}~\bibnamefont
  {Lehtinen}}, \bibinfo {author} {\bibfnamefont {B.}~\bibnamefont {Naydenov}},
  \bibinfo {author} {\bibfnamefont {P.}~\bibnamefont {B{\"{o}}rner}}, \bibinfo
  {author} {\bibfnamefont {K.}~\bibnamefont {Melentjevic}}, \bibinfo {author}
  {\bibfnamefont {C.}~\bibnamefont {M{\"{u}}ller}}, \bibinfo {author}
  {\bibfnamefont {L.~P.}\ \bibnamefont {McGuinness}}, \bibinfo {author}
  {\bibfnamefont {S.}~\bibnamefont {Pezzagna}}, \bibinfo {author}
  {\bibfnamefont {J.}~\bibnamefont {Meijer}}, \bibinfo {author} {\bibfnamefont
  {U.}~\bibnamefont {Kaiser}}, \ and\ \bibinfo {author} {\bibfnamefont
  {F.}~\bibnamefont {Jelezko}},\ }\href {\doibase 10.1103/PhysRevB.93.035202}
  {\bibfield  {journal} {\bibinfo  {journal} {Phys. Rev. B}\ }\textbf {\bibinfo
  {volume} {93}},\ \bibinfo {pages} {35202} (\bibinfo {year}
  {2016})}\BibitemShut {NoStop}%
\bibitem [{\citenamefont {Hanson}\ \emph {et~al.}(2008)\citenamefont {Hanson},
  \citenamefont {Dobrovitski}, \citenamefont {Feiguin}, \citenamefont {Gywat},\
  and\ \citenamefont {Awschalom}}]{Hanson2008}%
  \BibitemOpen
  \bibfield  {author} {\bibinfo {author} {\bibfnamefont {R.}~\bibnamefont
  {Hanson}}, \bibinfo {author} {\bibfnamefont {V.~V.}\ \bibnamefont
  {Dobrovitski}}, \bibinfo {author} {\bibfnamefont {a.~E.}\ \bibnamefont
  {Feiguin}}, \bibinfo {author} {\bibfnamefont {O.}~\bibnamefont {Gywat}}, \
  and\ \bibinfo {author} {\bibfnamefont {D.~D.}\ \bibnamefont {Awschalom}},\
  }\href {\doibase 10.1126/science.1155400} {\bibfield  {journal} {\bibinfo
  {journal} {Science}\ }\textbf {\bibinfo {volume} {320}},\ \bibinfo {pages}
  {352} (\bibinfo {year} {2008})}\BibitemShut {NoStop}%
\end{thebibliography}%
\clearpage

\end{document}